\documentclass[pdflatex,sn-mathphys-num]{sn-jnl}

\usepackage{graphicx}%
\usepackage{multirow}%
\usepackage{amsmath,amssymb,amsfonts}%
\usepackage{amsthm}%
\usepackage{mathrsfs}%
\usepackage[title]{appendix}%
\usepackage{xcolor}%
\usepackage[normalem]{ulem}%
\usepackage{textcomp}%
\usepackage{manyfoot}%
\usepackage{booktabs}%
\usepackage{algorithm}%
\usepackage{algorithmicx}%
\usepackage{algpseudocode}%
\usepackage{listings}%
\usepackage{subcaption}%
\usepackage{tabularx}%
\usepackage{CJKutf8}%

\graphicspath{{figures/}}
\usepackage[section]{placeins}

\usepackage{float}

\usepackage{derivative}

\usepackage{xr}
\makeatletter

\newcommand*{\addFileDependency}[1]{
\typeout{(#1)}

\@addtofilelist{#1}
\IfFileExists{#1}{}{\typeout{No file #1.}}
}\makeatother

\theoremstyle{thmstyleone}%
\theoremstyle{thmstyletwo}%

\theoremstyle{thmstylethree}%

\begin{document}

\title[FuXi-Nowcast]{Environment-conditioned deep learning for severe convection nowcasting}

\author[1]{\fnm{Lei}
\sur{Chen}}\email{cltpys@163.com}
\equalcont{These authors contributed equally to this work.}

\author[1]{\fnm{Zijian} \sur{Zhu}}\email{24210240443@m.fudan.edu.cn}
\equalcont{These authors contributed equally to this work.}

\author[2,3]{\fnm{Xiaoran} \sur{Zhuang}}\email{zxrxz3212009@163.com}
\equalcont{These authors contributed equally to this work.}

\author[1]{\fnm{Tianyuan} \sur{Qi}}\email{25213050312@m.fudan.edu.cn}

\author[2,3]{\fnm{Yuxuan} \sur{Feng}}\email{fengyuxuan27@163.com}
 
\author*[1,4]{\fnm{Xiaohui} \sur{Zhong}}\email{x7zhong@gmail.com}

\author*[1,4]{\fnm{Hao} \sur{Li}}\email{lihao$\_$lh@fudan.edu.cn}

\affil*[1]{\orgdiv{Artificial Intelligence Innovation and Incubation Institute}, \orgname{Fudan University}, \orgaddress{\city{Shanghai}, \postcode{200433}, \country{China}}}

\affil[2]{\orgname{Jiangsu Meteorological Observatory}, \orgaddress{\city{Nanjing, Jiangsu}, \postcode{210008}, \country{China}}}

\affil[3]{\orgname{Jiangsu Key Laboratory of Severe Storm Disaster Risk / Key Laboratory of Transportation Meteorology of CMA}, \orgaddress{\city{Nanjing, Jiangsu}, \postcode{210008}, \country{China}}}

\affil[4]{\orgname{FuXi Intelligent Computing Technology Co., Ltd.}, \orgaddress{\city{Shanghai}, \postcode{200233}, \country{China}}}


\abstract{Severe convection produces localized hazards that often require warnings before radar echoes fully reveal storm development. Convective initiation and the maintenance of intense convection remain challenging for radar-only nowcasting because pre-convective signals may be absent from recent radar observations and strong echoes often decay rapidly in forecasts. Here we present FuXi-Nowcast, an environment-conditioned deep learning system that combines high-resolution observations with three-dimensional atmospheric forecasts to predict composite reflectivity, precipitation, wind gusts, and surface variables up to 12 h ahead. In April--July 2024 evaluations over East China, FuXi-Nowcast outperforms operational numerical, persistence and extrapolation baselines for reflectivity and precipitation. Case studies, diagnostics, and ablation experiments suggest that atmospheric moisture information and explicit preservation of strong convective signals contribute to forecasts of convective initiation and maintenance. These results show that environmental conditioning can mitigate important failure modes of radar-only nowcasting for high-impact convective weather.}





\maketitle

\section{Introduction}

Severe convection often causes severe rainfall, damaging winds, and lightning into small, rapidly evolving systems, making it one of the most consequential tests of operational weather forecasting. High-impact weather in the first few forecast hours poses a major challenge for operational warning systems, particularly short-duration heavy rainfall and severe convective wind. Therefore, accurate high-resolution nowcasting is essential for disaster mitigation and timely decision-making \cite{raymond2020emergence,bojinski2023towards,tripathy2023climate,fowler2024precipitation, haslinger2025increasing}. Nowcasting, usually defined as short-range weather prediction over the next 0 to 6 hours \cite{schmid2019nowcasting, chen2020deep, franch2020precipitation}, relies on real-time, high-resolution observations to describe the current atmospheric state and its immediate evolution \cite{bojinski2023towards}. Conventional numerical weather prediction (NWP) systems often show limited skill at these lead times because of computational latency, infrequent updates, and model spin-up issues \cite{sun2014use, ravuri2021skilful, huang2012integrating}, especially for localized convective hazards \cite{sheridan2018current, majumdar2021multiscale, hess2022deep}.

A major challenge in short-range convective forecasting is convective initiation, the development of new convective echoes. It is defined as the first occurrence of radar reflectivity exceeding 35--40 dBZ \cite{Rita2003,bai2020convection}. Convective initiation is governed by complex interactions among moisture, instability, and dynamical lifting in the pre-convective environment \cite{Rita2003,bai2020convection}. These environmental signals cannot be inferred reliably from radar reflectivity alone. Classical methods such as TITAN typically identify and track echoes only after they have already reached 30--35 dBZ \cite{dixon1993titan}. Optical flow-based methods such as PySTEPS \cite{fleet2006optical,pulkkinen2019pysteps} extrapolate the displacement of existing echoes. However, such methods cannot predict convective initiation, subsequent intensification, or decay \cite{han20093d,agrawal2019machine,prudden2020review}. Many deep learning nowcasting models also mainly model the evolution of existing radar or precipitation fields and therefore still have limited ability to predict convective initiation.

Even after convection has formed, nowcasting systems often struggle to maintain the structure and intensity of severe convective systems. Consequently, extrapolation-based and radar-centered deep learning systems often lose skill after the first few hours, when storm growth, decay and organization become more important than echo advection \cite{han20093d,prudden2020review}. Many extrapolation-based and deep learning-based nowcasting methods smooth strong echoes, dissipate intense cells, or lose extreme precipitation signals as lead time increases \cite{ravuri2021skilful, zhang2023skilful, an2025deep}. These limitations are especially problematic for short-range heavy-rainfall forecasting. In that setting, the accurate representation of intense convective cores is closely related to warning value. Operational forecasting also requires more than precipitation alone. It must also address hazardous surface winds associated with severe convection \cite{ibrahim2020short, leclerc2025improving}. Therefore, a nowcasting framework that can better represent convective initiation and maintain intense convective signals remains needed.

Recent machine learning studies have advanced nowcasting along several related directions.
One group of methods focuses on deterministic evolution or extrapolation of radar and precipitation fields \cite{wang2017predrnn, ravuri2021skilful, zhang2023skilful}. A second group emphasizes generative, diffusion-based, or autoregressive refinement to improve spatial realism and extreme-event representation \cite{an2025deep, gong2024cascast, yu2024diffcast, gao2023prediff, franch2025gptcast}. Related developments have also expanded the nowcasting framework to other observation types and scales, including satellite-based thunderstorm nowcasting and convection-emulating forecast systems \cite{dai2024four, pathak2026kilometer}. A separate line of work extends forecast lead time by incorporating larger-scale atmospheric information, for example through the MetNet family and related systems \cite{sonderby2020metnet, espeholt2022deep, andrychowicz2023deep, agrawal2025operational}.
Despite these advances, most nowcasting methods still focus on the evolution of existing radar or precipitation fields, often for a single hazard, and rarely incorporate three-dimensional atmospheric fields to predict convective initiation.
Many also rely on NWP analyses or assimilation products, whose latency can limit real-time applicability in short-range warning operations.

Here we introduce FuXi-Nowcast, an environment-conditioned nowcasting framework designed to address two failure modes in severe convection prediction: missed initiation before radar echoes emerge and the rapid decay of intense signals during forecasting. The system conditions high-resolution prediction on three-dimensional atmospheric fields from FuXi-2.0 \cite{chen2023fuxi,zhong2024fuxi} (a deep learning-based weather forecasting model), together with radar, station, and High-Resolution China Meteorological Administration (CMA) Land Data Assimilation System (HRCLDAS) \cite{han2020evaluation} surface fields. 
To reduce intensity decay, the model combines Thresholded Signal Pooling (TSP), Adaptive Signal Fusion (ASF), and target-specific losses that emphasize strong convective structures during autoregressive prediction.
FuXi-Nowcast forecasts composite reflectivity, total precipitation, wind gusts, and surface variables up to 12 h ahead.
Results show that FuXi-Nowcast improves reflectivity and precipitation nowcasts relative to operational numerical, persistence and extrapolation baselines, with the clearest gains for convective initiation and maintenance of intense echoes over longer lead times.
Its low-latency and use of deep learning-based atmospheric guidance make it a practical alternative for operational severe weather nowcasting.
The model has also been deployed for operational nowcasting at the Jiangsu Meteorological Observatory (JMO).

\section{Results}

We evaluate the 12 h forecast performance of FuXi-Nowcast against three baselines: the China Meteorological Administration Mesoscale Model (CMA-MESO), Persistence, and PySTEPS, for total precipitation (TP), composite reflectivity (CR), and wind gust (GS), using data from April to July 2024, the period for which CMA-MESO baseline forecasts were available.
We first evaluate overall statistical performance and then analyze representative cases to assess convective initiation and maintenance.

\subsection{Overall statistical performance}

\begin{figure}[ht]
    \centering
    \includegraphics[width=\linewidth]{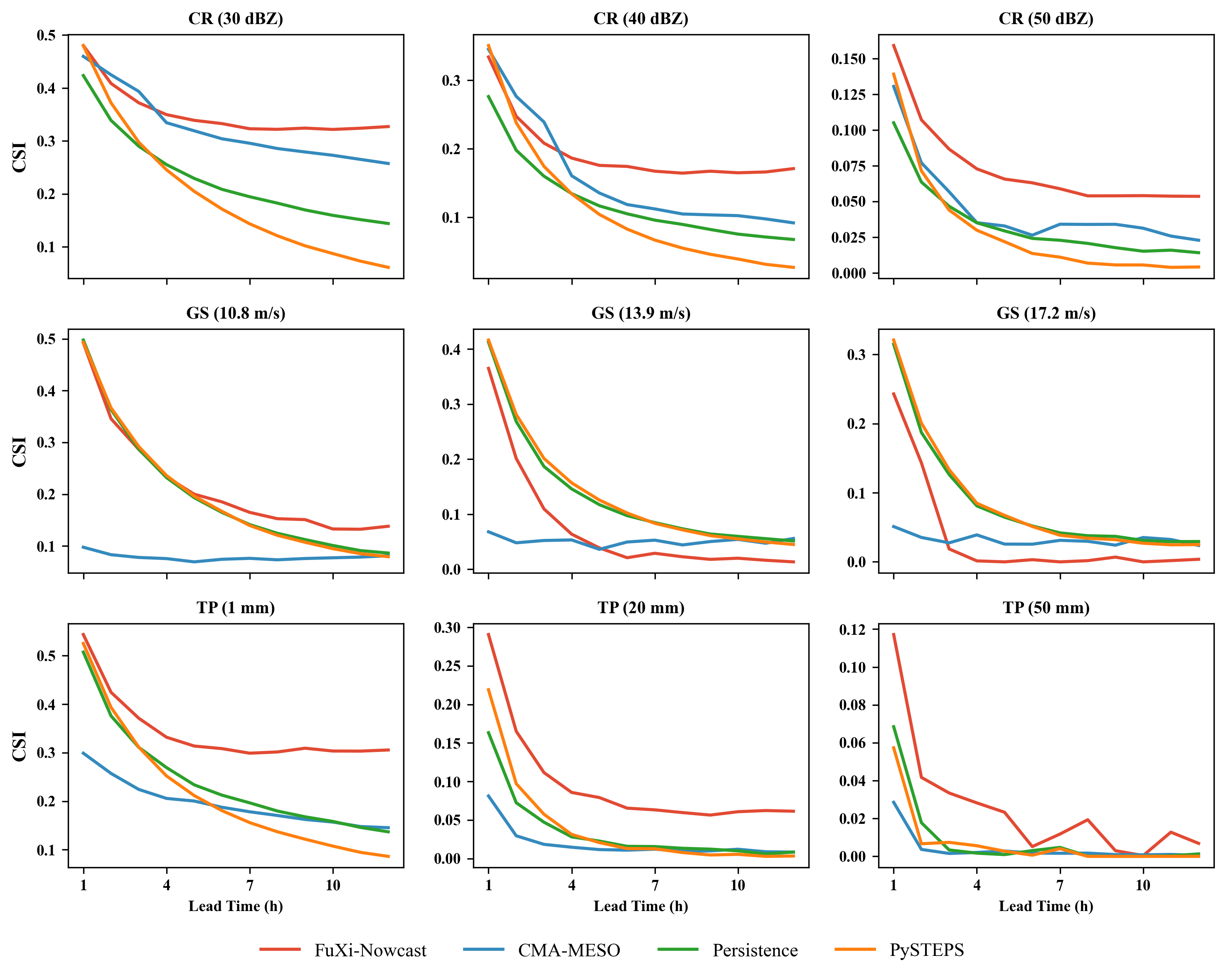}
    \caption{Critical success index (CSI) comparison among FuXi-Nowcast (red), CMA-MESO (blue), Persistence (green), and PySTEPS (orange) for 12 h forecasts. Rows correspond to composite reflectivity (CR), wind gust (GS), and total precipitation (TP), from top to bottom. Columns show the three intensity thresholds for each variable: CR at 30, 40, and 50 dBZ; GS at 10.8, 13.9, and 17.2 m/s; and TP at 1, 20, and 50 mm. Higher CSI indicates better forecast skill.}
    \label{fig:grapes_combined}
\end{figure}

We use the critical success index (CSI) as the primary metric, with frequency bias (BIAS), false alarm ratio (FAR), and probability of detection (POD) shown in Supplementary Figures \ref{fig:bias_comparison}--\ref{fig:pod_comparison}. CR and TP are evaluated on the full gridded domain using neighborhood verification, whereas GS is evaluated at station-selected grid points (Supplementary Figure \ref{Landsea_MASK_stations}). All forecasts are initialized every 3 h and verified at 1 h intervals.
Figure \ref{fig:grapes_combined} compares FuXi-Nowcast, CMA-MESO, Persistence, and PySTEPS for CR, GS, and TP at three thresholds. Skill generally decreases with lead time and event intensity.
For CR, FuXi-Nowcast performs best overall, especially at medium to longer lead times.
CMA-MESO is competitive only at short lead times for 30 and 40 dBZ, while Persistence and PySTEPS degrade rapidly.
At 50 dBZ, FuXi-Nowcast maintains the highest CSI throughout the forecast period.

The largest differences occur for GS. CMA-MESO has near-zero CSI across thresholds. FuXi-Nowcast becomes the best-performing model after the first few hours at 10.8 m/s, but Persistence and PySTEPS remain superior at 13.9 and 17.2 m/s. For TP, FuXi-Nowcast outperforms all baselines at 1 mm and remains the best model at 20 and 50 mm, although CSI is low for all methods. Overall, FuXi-Nowcast shows its clearest advantage for CR and TP, with the CR advantage increasing at longer lead times.

\FloatBarrier

\subsection{Convective event forecasting}

We first present a representative dryline-triggered convective initiation case to illustrate the comparative performance of FuXi-Nowcast, CMA-MESO, and PySTEPS. As the statistical evaluation uses April--July 2024 data, we treat the 2025 events as independent qualitative case studies rather than as part of the benchmark evaluation.. We then present a complementary case that focuses on the maintenance of organized convection. Additional convective event cases are shown in Supplementary Figures \ref{fig:20240612T03} and \ref{fig:20240621T06}, and Supplementary Figure \ref{fig:case2_combined} shows the environmental diagnostics for the 27 June 2025 squall-line case.

\subsubsection{Dryline-triggered convective initiation: 16 June 2025}

\begin{figure}[ht]
    \centering
    \includegraphics[width=\linewidth]{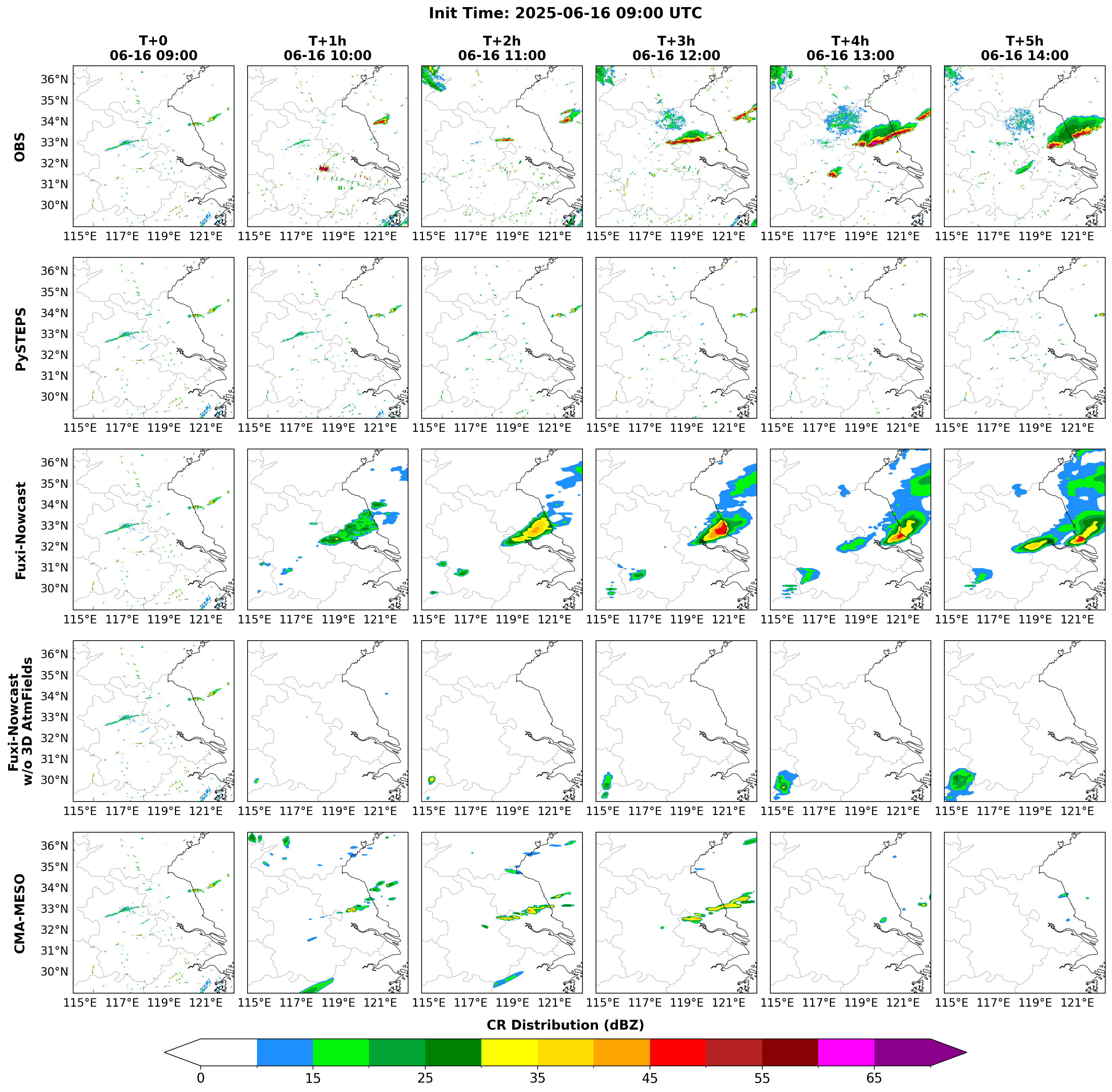}
    \caption{Composite radar reflectivity (dBZ) over East China (29.01\textdegree N -- 36.68\textdegree N, 114.67\textdegree E -- 122.34\textdegree E) for a dryline-triggered convective initiation case initialized at 09 UTC 16 June 2025. Rows from top to bottom show observations, PySTEPS forecasts, FuXi-Nowcast forecasts, the retrained FuXi-Nowcast variant without the three-dimensional atmospheric fields, and CMA-MESO forecasts, respectively. Columns from left to right correspond to forecast lead times of 0 to 5 hours, valid from 09 to 14 UTC. The color bar denotes reflectivity values from 0 to 65 dBZ.}
    \label{fig:20250616T09}
\end{figure}

A dryline-triggered convective event occurred over East China on 16 June 2025. Drylines are well-known convective-initiation boundaries \cite{trier2015mesoscale,ziegler1997convective,ziegler1998initiation}, so this case provides a useful test of whether the model can use pre-convective environmental signals to anticipate convective initiation.
As shown in Figure \ref{fig:20250616T09}, observed CR shows scattered, mostly weak echoes at the initialization time (09 UTC) and at T + 1 h (10 UTC).
By T + 2 h (11 UTC), new convective echoes begin to emerge in the central part of East China, and by T + 3 h (12 UTC) a well-organized linear convective system develops with reflectivity exceeding 50 dBZ.
The convection continues to intensify and expand through T + 4--5 h (13--14 UTC), forming a broad band across the central and eastern parts of East China.

FuXi-Nowcast captures the main evolution of this convective initiation sequence. At T + 1 h, it predicts weak echoes below the 35 dBZ convective initiation threshold. By T + 2 h, organized convection emerges in the central part of East China. The forecast then intensifies into a linear structure through T + 3--5 h. The predicted spatial distribution, orientation, and intensity evolution are generally consistent with the radar observations.
PySTEPS extrapolates the weak initial echoes and produces very sparse fields throughout the forecast period. This behavior is consistent with the difficulty of predicting convective initiation when the pre-existing radar signal is weak. CMA-MESO produces some scattered echoes over the central part of East China. However, these echoes do not continue to develop into the observed linear convective system. Instead, they dissipate too early after T + 2--3 h.
The retrained variant without the three-dimensional atmospheric fields shows a much weaker response: it retains only limited echoes near the southwestern corner of the domain and fails to develop the organized convective band seen after T + 2 h.

This case suggests that FuXi-Nowcast can predict this convective initiation and subsequent convective evolution more realistically than the baselines and the variant without atmospheric fields. It also indicates that the three-dimensional atmospheric fields provide guidance that is not available from observation-based extrapolation alone. The contrast between the full model and the retrained variant without atmospheric fields provides case-based evidence that these fields play an important role in convective initiation prediction, although broader convective initiation-specific verification is still needed.

\FloatBarrier

\subsubsection{Maintenance of organized convection: 14 June 2024}

\begin{figure}[ht]
    \centering
    \includegraphics[width=0.98\linewidth]{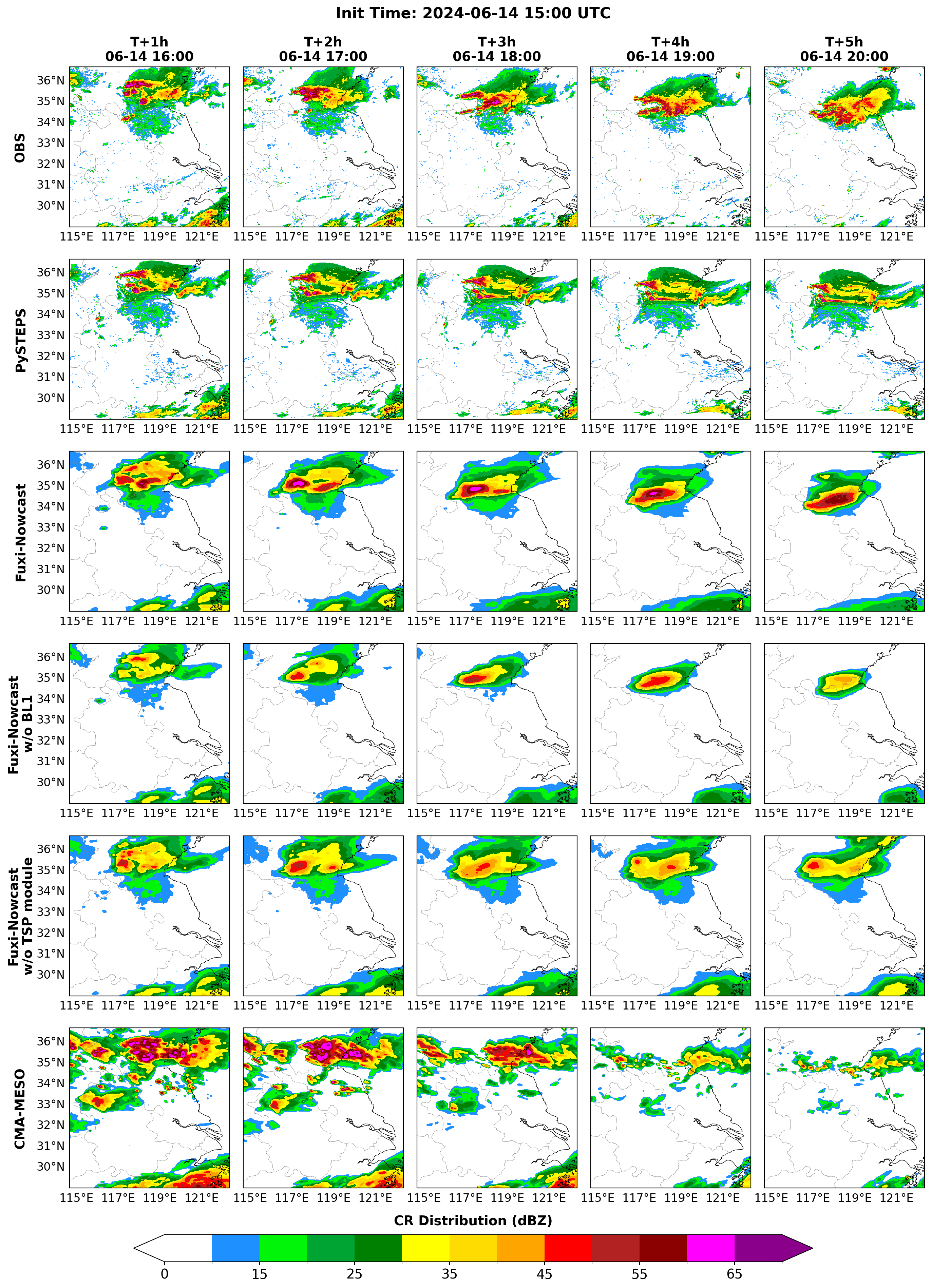}
    \caption{Composite radar reflectivity (dBZ) for the organized-convection maintenance case initialized at 15 UTC 14 June 2024. Columns from left to right correspond to forecast lead times of T + 1 to T + 5 h, valid from 16 to 20 UTC. Rows show observations, PySTEPS, FuXi-Nowcast, two retrained ablation variants without the Thresholded Signal Pooling module (TSP module) and without the Balanced L1 loss (BL1), and CMA-MESO. The comparison highlights that the full FuXi-Nowcast better maintains the spatial organization and intensity of the convective band over time.}
    \label{fig:20240614T15_main}
\end{figure}

To complement the convective initiation-focused dryline case, we also examine the strong-convection event of 14 June 2024, which emphasizes the maintenance of organized convection. As shown in Figure \ref{fig:20240614T15_main}, observed CR exhibits an organized convective band over the northern part of East China from T + 1 h (16 UTC) to T + 5 h (20 UTC), with persistent intense echoes exceeding 50 dBZ. FuXi-Nowcast reproduces both the eastward propagation and the sustained high-intensity structure of this system more realistically than PySTEPS and CMA-MESO. PySTEPS largely preserves the initial echo pattern but does not capture the observed structural evolution of the convective band as lead time increases, whereas CMA-MESO produces overly strong and spatially fragmented echoes at early lead times and then dissipates them too rapidly.
Relative to the full FuXi-Nowcast, the variants without the Thresholded Signal Pooling module (TSP module) or without the Balanced L1 loss (BL1) weaken the structure and intensity of the convective band more rapidly as lead time increases. This comparison provides a case-based illustration that the TSP module and BL1 help preserve strong convective signals during autoregressive prediction.


\FloatBarrier

\subsection{Analyses of physical mechanisms}

\begin{figure}[ht]
    \centering
    \includegraphics[width=\linewidth]{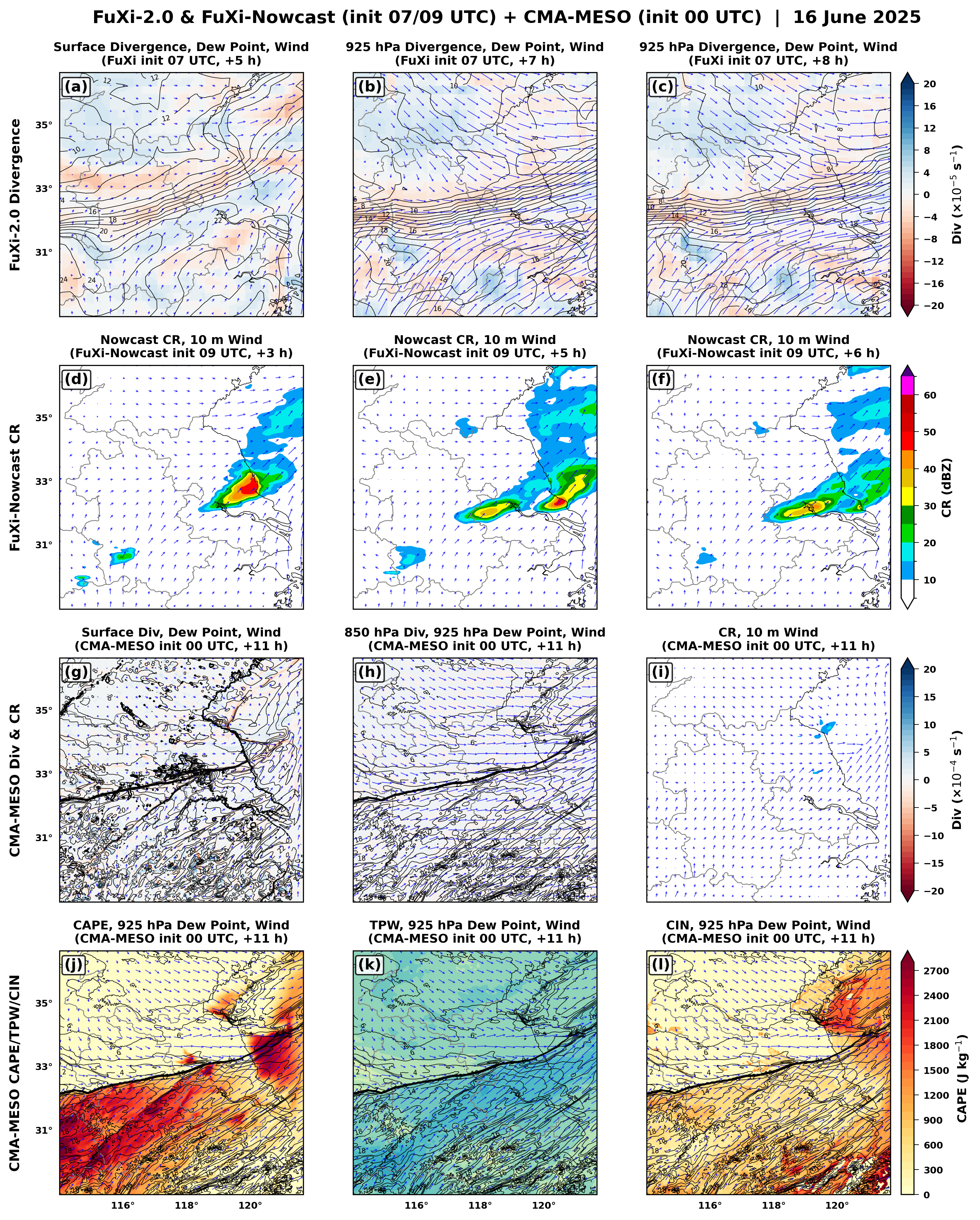}
    \caption{Combined diagnostic figure for the dryline-triggered convective initiation case on 16 June 2025. Panels (a)--(c) show offline environmental diagnostics derived from the FuXi-2.0 atmospheric fields, including divergence (shading, $\times 10^{-5}$~s$^{-1}$), dew point temperature (contours, $^{\circ}$C), and wind vectors, at the surface for valid 12 UTC and at 925 hPa for valid 14 and 15 UTC. Panels (d)--(f) show the corresponding FuXi-Nowcast reflectivity forecasts (shading, dBZ) and 10-m wind vectors for valid 12, 14, and 15 UTC. Panels (g)--(l) show CMA-MESO forecast diagnostics valid at 11 UTC, including surface divergence, 850 hPa divergence, composite reflectivity, CAPE, TPW, and CIN, all overlaid with dew point contours and wind vectors as indicated in the panel titles.}
    \label{fig:combined_diagnostic}
\end{figure}

To diagnose the environmental context associated with this successful initiation forecast, we analyze mesoscale fields derived from FuXi-2.0. Figure \ref{fig:combined_diagnostic} combines the offline environmental diagnostics derived from the FuXi-2.0 atmospheric fields, the corresponding FuXi-Nowcast reflectivity forecasts, and representative CMA-MESO forecast diagnostics. We then compare these diagnostics across the FuXi-Nowcast and CMA-MESO forecasts. The diagnostic fields shown here, including dew-point temperature, wind divergence, convective available potential energy (CAPE), total precipitable water (TPW), and convective inhibition (CIN), are not direct inputs to FuXi-Nowcast. They are computed offline from the atmospheric fields and are used here only as indicators of the atmospheric environment.

At 12 UTC, the FuXi-2.0 surface fields show a clear dew point gradient near the central East China coastline (Figure \ref{fig:combined_diagnostic}a). This feature is consistent with a dryline-like boundary formed by the confluence of northwesterly and southwesterly winds. A distinct convergence center develops near the coastline along this boundary. The predicted convective core is collocated with this convergence center (Figure \ref{fig:combined_diagnostic}d). As the system evolves, the convergence gradually deepens from the surface to 925 hPa. By 14 UTC, the 925 hPa convergence zone shifts to the northern side of this dryline-like boundary near the coast (Figure \ref{fig:combined_diagnostic}b). New convection then develops near the coastline (Figure \ref{fig:combined_diagnostic}e). By 15 UTC, the low-level convergence remains evident near the coast (Figure \ref{fig:combined_diagnostic}c), while the corresponding reflectivity forecast still shows organized convection there (Figure \ref{fig:combined_diagnostic}f). Overall, the atmospheric fields represent a coherent sequence of convergence along this dryline-like boundary, upward deepening of the convergence layer, and subsequent coastal convective development.

In the CMA-MESO panels of Figure \ref{fig:combined_diagnostic}, CMA-MESO also produces a dryline across the central part of East China, but the feature is overly narrow and the associated surface convergence line is unrealistically thin (Figure \ref{fig:combined_diagnostic}g). Weak convective echoes briefly appear near the coast at 11 UTC (Figure \ref{fig:combined_diagnostic}i), but they do not develop further. At 850 hPa, a convergence line collocated with the dryline indicates that the dynamical lifting extends through a substantial depth (Figure \ref{fig:combined_diagnostic}h). The thermodynamic environment on the moist side of the dryline also appears favorable. CAPE exceeds 2500 J\,kg$^{-1}$ (Figure \ref{fig:combined_diagnostic}j), and total precipitable water exceeds 50 kg\,m$^{-2}$ (Figure \ref{fig:combined_diagnostic}k). Even with these ingredients, convection does not develop further in the CMA-MESO forecast. Moderately elevated CIN on the moist side of the dryline may contribute to this behavior (Figure \ref{fig:combined_diagnostic}l). However, the immediate reason for the forecast failure cannot be isolated from this single case alone.

Together, the diagnostics and ablation experiments support the interpretation that three-dimensional moisture and dynamical information helps the model to predict convective initiation in this case. Consistent with this interpretation, the retrained variant without the three-dimensional atmospheric fields in Figure \ref{fig:20250616T09} retains only limited echoes near the southwestern corner of the domain and fails to develop the organized post-initiation convection captured by the full model. The ablation experiments support the same interpretation: among all input variable groups, removing specific humidity causes the largest degradation in CR forecast skill (Supplementary Figures \ref{fig:ablation_combined} and \ref{fig:method_ablation_csi}). This result indicates that moisture-related atmospheric fields are particularly important for the model's statistical reflectivity skill.

\FloatBarrier

\subsection{Effect of the atmospheric field source}

FuXi-Nowcast is trained with ERA5 reanalysis but uses FuXi-2.0 forecasts during inference.
This design allows the model to learn from physically consistent reanalysis while retaining the ability to ingest real-time forecast fields.
To quantify the effect of the atmospheric field source, we compare FuXi-Nowcast inference driven by ERA5 and FuXi-2.0 using identical model weights.
As ERA5 is not available in real time at the forecast target time, the ERA5 experiment is not an operational inference configuration and is used only to diagnose the atmospheric field gap.
FuXi-2.0 is used for operational inference because its low latency and hourly forecasts are well matched to the nowcasting framework.
This comparison quantifies a training–inference source shift.
Table \ref{tab:era5_vs_fuxi} summarizes the CSI comparison for representative variables, thresholds, and lead times.

The source impact depends on the variable, threshold, and lead time.
CR differences are small at short lead times but increase at longer lead times, especially for higher thresholds. For TP, FuXi-2.0 improves the 20 mm threshold at short lead times but degrades longer-lead performance, particularly at 1 mm. For GS, FuXi-2.0 matches or exceeds ERA5 at 10.8 m/s but performs worse at 17.2 m/s. The largest relative degradations occur for CR at 50 dBZ at T + 12 h ($-$25.4\%), TP at 1 mm at T + 12 h ($-$24.4\%), and GS at 17.2 m/s at T + 6 h ($-$85.7\%). FuXi-2.0 improves TP at 20 mm at T + 1 h (+28.0\%) and GS at 10.8 m/s at T + 12 h (+66.3\%).
Thus, the source gap is not uniform, and FuXi-2.0 remains the practical atmospheric field source for operational inference because it is available with low latency and is skillful in key nowcasting targets.

\begin{table}[ht]
    \centering
        \caption{CSI comparison between ERA5 reanalysis and FuXi-2.0 forecasts as three-dimensional atmospheric fields during inference.}
    \scriptsize
    \setlength{\tabcolsep}{5pt}
    \begin{tabular}{@{}ll*{4}{cc}@{}}
        \toprule
        & & \multicolumn{2}{c}{\textbf{T + 1 h}} & \multicolumn{2}{c}{\textbf{T + 3 h}} & \multicolumn{2}{c}{\textbf{T + 6 h}} & \multicolumn{2}{c}{\textbf{T + 12 h}} \\
        \cmidrule(lr){3-4} \cmidrule(lr){5-6} \cmidrule(lr){7-8} \cmidrule(lr){9-10}
        \textbf{Variable} & \textbf{Threshold} & ERA5 & FuXi & ERA5 & FuXi & ERA5 & FuXi & ERA5 & FuXi \\
        \midrule
        CR & 30 dBZ & 0.470 & 0.479 & 0.371 & 0.367 & 0.336 & 0.309 & 0.328 & 0.296 \\
        CR & 50 dBZ & 0.157 & 0.168 & 0.086 & 0.082 & 0.064 & 0.061 & 0.059 & 0.044 \\
        TP & 1 mm   & 0.423 & 0.490 & 0.338 & 0.312 & 0.297 & 0.253 & 0.295 & 0.223 \\
        TP & 20 mm  & 0.268 & 0.343 & 0.103 & 0.122 & 0.062 & 0.057 & 0.056 & 0.043 \\
        GS & 10.8 m/s & 0.492 & 0.492 & 0.257 & 0.286 & 0.151 & 0.185 & 0.083 & 0.138 \\
        GS & 17.2 m/s & 0.313 & 0.243 & 0.073 & 0.019 & 0.021 & 0.003 & 0.008 & 0.004 \\
        \bottomrule
    \end{tabular}
    \label{tab:era5_vs_fuxi}
\end{table}
\FloatBarrier

\section{Discussion}

FuXi-Nowcast shows that severe convection nowcasting can be formulated as an environment-conditioned and signal preserving sequence prediction problem. Radar-only nowcasting methods face two major limitations: they cannot infer new convective development when precursors are absent from the target reflectivity field, and they often smooth or dissipate intense convective signals during autoregressive prediction. By conditioning high-resolution forecasts on three-dimensional atmospheric fields and explicitly enhancing strong convection signals, FuXi-Nowcast mitigates both failure modes. This formulation provides a deep learning strategy for forecasting high-impact events whose precursors are weak in the target observation stream and whose extremes are easily lost during sequential prediction.

The case analyses and diagnostics indicate that this improvement is linked to environmental signals in the atmospheric fields. In the dryline-triggered case, predicted convective initiation occurs near low-level convergence and strong thermodynamic gradients (Figures \ref{fig:20250616T09} and \ref{fig:combined_diagnostic}). By contrast, the retrained variant without three-dimensional atmospheric fields retains only limited echoes near the southwestern domain and fails to reproduce the subsequent organized convection (Figure \ref{fig:20250616T09}), supporting the importance of atmospheric context for convective initiation prediction. Ablations further show that removing specific humidity causes the largest degradation in reflectivity skill (Supplementary Figures \ref{fig:ablation_combined} and \ref{fig:method_ablation_csi}), indicating the importance of moisture-related fields, although this result does not isolate their role in convective initiation.

FuXi-Nowcast also better preserves strong convective intensity at longer lead times, reflecting both the TSP and ASF modules and the loss design. In the 14 June 2024 case (Figure \ref{fig:20240614T15_main}), the full model maintains the organized convective band more effectively than variants without the TSP module or BL1. Component ablations confirm that removing the TSP module reduces strong-echo skill, whereas removing Balanced L1 markedly degrades heavy-precipitation skill (Supplementary Figure \ref{fig:method_ablation_csi}). Thus, environmental guidance alone is insufficient and targeted treatment of intense events is also needed for high-resolution nowcasting.

Despite the promising results, several limitations remain. First, convective initiation localization is still imperfect, with case-study position errors of tens of kilometers, partly because coarse atmospheric fields limit the representation of mesoscale boundaries such as drylines and sea-breeze fronts. Second, evaluation is limited to East China, so performance in other climatic and convective regimes remains untested. Third, the 1 h temporal resolution may be insufficient for rapidly evolving convective life cycles. Fourth, the deterministic system does not provide forecast uncertainty, which is important for operational warnings. Finally, broader application requires high-quality multi-source meteorological datasets with consistent coverage, resolution, and quality control, which remain difficult to assemble even within a single region.

Overall, these results suggest that severe convection nowcasting benefits from treating initiation and maintenance as complementary prediction challenges: one requiring environmental context before radar signals emerge, and the other requiring explicit preservation of intense signals during autoregressive prediction. Future work should improve spatial localization using finer-resolution atmospheric predictors, extend the deterministic framework to probabilistic forecasting \cite{zhong2025fuxi}, and test transferability across broader climatic regimes through regional adaptation with heterogeneous local observation networks.
  
\section{Methods}

FuXi-Nowcast is designed as a conditional autoregressive forecasting model designed around two requirements for severe-convection nowcasting: environmental awareness before initiation and signal preservation after intense echoes. The first component supplies three-dimensional atmospheric context that is not directly available from the recent history of radar and surface fields. The second component enhances high-intensity convective signals and uses target-specific losses to reduce smoothing and intensity decay during autoregressive prediction.

\subsection{Data}

FuXi-Nowcast uses ERA5 reanalysis, FuXi-2.0 forecasts, radar composite reflectivity, observations from ground-based weather stations, HRCLDAS, and a static land-sea mask. Supplementary Table \ref{tab:all_data_si} summarizes their variables, roles, and spatiotemporal resolutions. ERA5 provides the three-dimensional atmospheric fields during training, whereas the most recent FuXi-2.0 forecasts replace ERA5 during inference to support real-time application. The model is trained with April--September data from 2019--2023, evaluated on April--July 2024, and further tested using selected 2025 cases. Additional details on data sources, preprocessing, the station network, normalization, and training-sample filtering are provided in Supplementary Table \ref{tab:all_data_si}, Supplementary Figure \ref{Landsea_MASK_stations}, and Supplementary Figures \ref{fig:variable_distributions} and \ref{fig:variables_distribution_filter}.

\subsection{CMA-MESO 3-km model}

The outputs from the China Meteorological Administration Mesoscale Model (CMA-MESO) are used in this study.
Specifically, the 3-km resolution configuration (CMA-MESO 3-km) serves as the baseline NWP model for comparison with FuXi-Nowcast.
This operational system, driven by initial and boundary conditions from the global CMA-GFS model, provides forecasts covering China and its surrounding regions \cite{liping2022key}.
For evaluation, we use CMA-MESO 3-km outputs of TP and diagnosed CR.
Because CMA-MESO does not provide native GS output, we multiply the 10-m wind speed by a factor of 1.75 to derive GS.
The resulting GS scores should be interpreted as a diagnosed CMA-MESO baseline rather than as native CMA-MESO GS output.
All 3-km CMA-MESO forecasts are interpolated to a 1-km grid over the target domain.

\subsection{FuXi-Nowcast model}

\begin{figure}[ht]
    \centering
    \includegraphics[width=\linewidth]{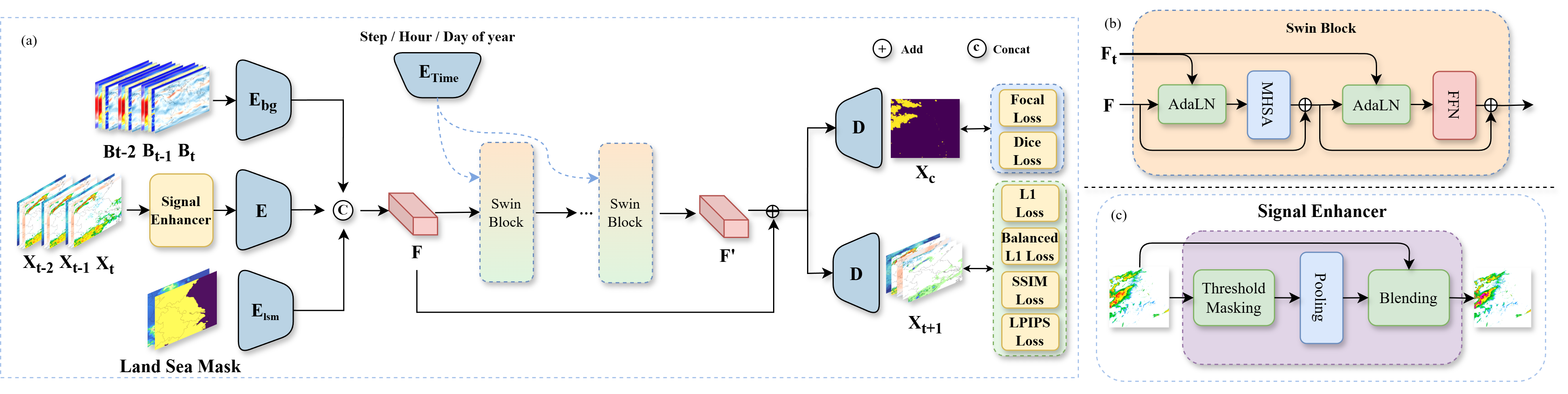}
    \caption{(a) Architecture of FuXi-Nowcast, an environment-conditioned and signal-preserving autoregressive nowcasting framework. The model takes two groups of inputs: FuXi-2.0 three-dimensional atmospheric fields ($B_{t-2}$, $B_{t-1}$, $B_t$) encoded by $E_{\mathrm{bg}}$, and high-resolution nowcast fields ($X_{t-2}$, $X_{t-1}$, $X_t$) first processed by the TSP and ASF modules and then encoded by $E$. A land-sea mask is encoded by $E_{\mathrm{lsm}}$, and temporal features (forecast step, hour, and day of year) are encoded by $E_{\mathrm{Time}}$. The encoded features are combined via concatenation~($\copyright$), then passed through a sequence of Swin Blocks. The decoder has two branches: a primary forecasting branch~(D) producing the next-step prediction $X_{t+1}$, and an auxiliary precipitation classification branch producing classification output $X_c$. The model is trained with a hybrid loss comprising Focal, Dice, L1, Balanced L1, SSIM, and LPIPS losses. (b) Internal structure of the Swin Block. The temporal feature $F_t$ is injected via adaptive layer normalization (AdaLN) before the multi-head self-attention (MSA) and feed-forward network (FFN) layers, with residual connections ($\oplus$) at each stage. (c) Design of the TSP and ASF modules, which apply threshold masking, pooling, and adaptive fusion to selectively reinforce strong convective signals.}
    \label{fig:nowcast_structure}    
\end{figure}

FuXi-Nowcast addresses convective initiation, intensity decay, and multi-hazard prediction by combining FuXi-2.0 three-dimensional atmospheric guidance, high-resolution local observations, the TSP and ASF modules, and a hybrid training objective with an auxiliary TP classification task. The architecture is shown in Figure \ref{fig:nowcast_structure}.

The model predicts autoregressively from an input tensor with shape $\textrm{B} \times \textrm{T} \times \textrm{C} \times \textrm{H} \times \textrm{W}$, where $\textrm{B}$ is the batch size, $\textrm{T}$ is the number of preceding time steps, $\textrm{C}$ is the number of input variables, and $\textrm{H}$ and $\textrm{W}$ are the latitude and longitude grid sizes, respectively. We use $\textrm{T}=3$ time steps ($t-2$, $t-1$, and $t$), $\textrm{C}=78$ variables, including 8 high-resolution local variables and 70 atmospheric-field variables, and $\textrm{H}=\textrm{W}=768$. Atmospheric fields are from ERA5 during training and FuXi-2.0 during inference.

Before embedding, the TSP module reinforces strong WS, GS, CR, and TP signals with a kernel size of 15 and normalized thresholds of 0.31, 0.27, 0.57, and 0.05, corresponding to 10.8 m/s, 10.8 m/s, 40 dBZ, and 5 mm. These thresholds are selected through the ablation experiments summarized in Supplementary Table \ref{tab:ablation_settings} and Supplementary Figures \ref{fig:ablation_combined} and \ref{fig:method_ablation_csi}. The ASF module then uses channel-wise learnable alpha parameters to control how much of the enhanced signal is mixed back into the original input. An alpha value of 0 retains the original input, whereas larger values increase the contribution of the enhanced signal. The alpha parameters are initialized as T2M = 10, Q2M = 10, U10M = 10, V10M = 10, WS = 0, GS = 0, CR = 10, and TP = 10. The enhanced variables, static fields, and temporal features are encoded separately and then combined.

The feature extractor contains 24 Swin Transformer blocks with $96 \times 96$ windows and 24 attention heads. The decoder restores the latent feature map from $\textrm{B} \times 1024 \times \textrm{H}/8 \times \textrm{W}/8$ to the original $768 \times 768$ grid. Its primary branch predicts the eight target variables, while an auxiliary branch predicts 11 TP classes defined by 10 intensity thresholds (0.5, 1, 2, 5, 10, 15, 20, 30, 40, and 50 mm). The auxiliary branch is used only during training. Each predicted field is appended to the historical sequence and combined with the corresponding FuXi-2.0 atmospheric variables for the next autoregressive step, enabling forecasts up to 12 h.

\FloatBarrier

\subsection{FuXi-Nowcast model training}

The target variables have different distributions before and after the training-sample filtering procedure (Supplementary Figures \ref{fig:variable_distributions} and \ref{fig:variables_distribution_filter}): T2M, U10M, and V10M are nearly symmetric, whereas TP and CR are sparse, skewed, and long-tailed. We therefore use a hybrid loss $\textrm{L}_\textrm{total}$ with target-specific regression, perceptual, structural, and classification terms:

\begin{equation}
\begin{split}
    \mathrm{L}_{\mathrm{total}} ={}& \lambda_{\mathrm{cl1}} \mathrm{L}_{\mathrm{cl1}} + \lambda_{\mathrm{bl1}} \mathrm{L}_{\mathrm{bl1}} + \lambda_{\mathrm{lpips}} \mathrm{L}_{\mathrm{lpips}} \\
    & + \lambda_{\mathrm{SSIM}} \mathrm{L}_{\mathrm{SSIM}} + \lambda_{\mathrm{focal}} \mathrm{L}_{\mathrm{focal}} + \lambda_{\mathrm{dice}} \mathrm{L}_{\mathrm{dice}}
\end{split}
\end{equation}

Here, $\mathrm{L}_{\mathrm{cl1}}$, $\mathrm{L}_{\mathrm{bl1}}$, $\mathrm{L}_{\mathrm{lpips}}$, $\mathrm{L}_{\mathrm{SSIM}}$, $\mathrm{L}_{\mathrm{focal}}$, and $\mathrm{L}_{\mathrm{dice}}$ denote Charbonnier, Balanced L1, LPIPS, SSIM, Focal, and Dice losses, with weights 1.0, 0.2, 0.1, 0.05, 0.1, and 0.1, respectively.

The Charbonnier loss \cite{charbonnier1994two} is applied to all eight output variables:
\begin{equation}
\label{eq:charbonnier_loss}
\mathrm{L}_\mathrm{cl1}=\sqrt{(\hat{\mathbf{X}} - \mathbf{X})^2+\epsilon^2}
\end{equation}
where $\mathbf{X}$ is the ground truth, $\hat{\mathbf{X}}$ is the prediction, and $\epsilon=10^{-6}$.

For relatively rare high-intensity TP, the Balanced L1 loss \cite{pang2019libra, shi2017deep} assigns larger weights to less frequent bins:

\begin{equation}
\label{eq:bl1_loss}
\mathrm{L}_{\mathrm{bl1}}=
\frac{\sum_{i=1}^{N} w(\mathbf{X}_i)\,|\mathbf{X}_i-\hat{\mathbf{X}}_i|}
{\sum_{i=1}^{N} w(\mathbf{X}_i)+\epsilon}
\end{equation}


where $N$ is the number of TP target elements included in the loss computation, and $w(\mathbf{X}_i)=w_k$ if $\mathbf{X}_i$ falls into bin $k$. The bin weight is defined as $w_k=(1/f_k)^\gamma$, where $f_k$ is the exponentially averaged frequency of bin $k$. This $\mathrm{L}_\mathrm{bl1}$ term is applied only to TP.

To reduce blurring, we use the LPIPS loss $\mathrm{L}_{\mathrm{lpips}}$ with a pre-trained VGG network:

\begin{equation}
\label{eq:lpips_loss}
\mathrm{L}_\mathrm{lpips}=
\sum_l\frac{1}{\mathrm{H}_l\mathrm{W}_l} \sum_{h,w}
\left\| w_l \odot \left(f_l(\mathbf{X})_{hw}-f_l(\hat{\mathbf{X}})_{hw}\right) \right\|_2^2
\end{equation}

where $f_l(\cdot)_{hw}$ denotes the feature vector at spatial location $(h,w)$ in layer $l$ of the pre-trained VGG network, $w_l$ denotes the learned channel-wise weights in layer $l$, and $\odot$ denotes element-wise multiplication. This term is applied to WS, GS, CR, and TP.

We add SSIM loss to encourage structural consistency:
\begin{equation}
    \mathrm{L}_{\text{SSIM}} = 1 - \frac{(2\mu_\mathrm{X}\mu_{\hat{\mathrm{X}}} + c_1)(2\sigma_{\mathrm{X}\hat{\mathrm{X}}} + c_2)}{(\mu_{\mathrm{X}}^2 + \mu_{\hat{\mathrm{X}}}^2 + c_1)(\sigma_{\mathrm{X}}^2 + \sigma_{\hat{\mathrm{X}}}^2 + c_2)}
\end{equation}
where $\mu$, $\sigma^2$, and $\sigma_{\mathrm{X\hat{X}}}$ denote image mean, variance, and covariance, with $c_1=c_2=10^{-6}$. This term is also applied to WS, GS, CR, and TP.

For TP classification, we use Focal loss to address the imbalance between non-raining and raining pixels:
\begin{equation}
    \mathrm{L}_{\text{focal}} = -\alpha_t(1-p_t)^\gamma \log(p_t)
\end{equation}
where $p_t$ is the predicted probability of the true class. In our implementation, $\gamma = 1.5$ and $\alpha_t = 0.25$.

Finally, Dice loss improves precipitation-region overlap:

\begin{equation}
    \mathrm{L}_{\text{dice}} = 1 - \frac{2|\mathbf{X} \cap \hat{\mathbf{X}}| + \epsilon}{|\mathbf{X}| + |\hat{\mathbf{X}}| + \epsilon}
\end{equation}

where $|\mathrm{\mathbf{X} \cap \hat{\mathbf{X}}}|$ is the shared predicted--observed region and $\epsilon=10^{-6}$. This term is applied to TP classification.

The model is implemented in PyTorch \cite{paszke2019pytorch} and optimized end-to-end with AdamW \cite{kingma2014adam, loshchilov2017decoupled} ($\beta_1=0.9$, $\beta_2=0.95$, weight decay 0.1). The learning rate warms up linearly for the first 2\% of steps to 2.5$\times10^{-4}$ and then decays cosine-wise to $5\times10^{-6}$. Training uses two NVIDIA A100 GPUs, a per-GPU batch size of 1, gradient accumulation over two steps, and 30,000 iterations. Training takes approximately 24 h on two NVIDIA A100 GPUs, and one 12 h autoregressive forecast takes about 10 s on one NVIDIA A100 GPU.

\subsection{Evaluation method}

The main verification metric is the critical success index (CSI), with BIAS, FAR, and POD shown in Supplementary Figures \ref{fig:bias_comparison}--\ref{fig:pod_comparison}. CR and TP are evaluated on the full gridded domain using neighborhood-based spatial verification, whereas GS is evaluated at grid points selected by stations. The metric definitions and neighborhood procedure are provided in the supplementary section ``Evaluation methods'', and the diagnostic-variable derivations are provided in the supplementary section ``Diagnostic variables''.

\section*{Data availability}

The ERA5 reanalysis data are accessible through the Copernicus Climate Data Store at \url{https://cds.climate.copernicus.eu/}.
ECMWF HRES forecasts can be retrieved from \url{https://apps.ecmwf.int/archive-catalogue/?type=fc&class=od&stream=oper&expver=1}.
The data that support the findings of this study are available from Jiangsu Meteorological Observatory (JMO) of China Meteorological Administration (CMA).
Restrictions apply to the availability of these data, which were used under license for this study.

\section*{Code availability}

The source scripts used for running the FuXi-Nowcast model are available from Zenodo at \url{https://doi.org/10.5281/zenodo.19911638} \cite{nowcast_code}.

\section*{Acknowledgements}
This work was supported by the AI for Science Program, Shanghai Municipal Commission of Economy and Informatization (2025-GZL-RGZN-BTBX-02017), and the National Natural Science Foundation of China (NSFC) under Grant 42505143.
We extend our sincere appreciation to the researchers at ECMWF for their invaluable contributions in collecting, archiving, disseminating, and maintaining the ERA5 reanalysis and HRES.
We would like to thank the China Meteorological Administration for the HRCLDAS data.
We also thank the Jiangsu Meteorological Observatory for providing the radar composite reflectivity and ground weather station data used in this study.

The computations in this research were performed using the CFFF platform of Fudan University.

\section*{Competing interests}
The authors declare no competing interests.

\noindent





\clearpage
\section*{Supplementary Information}
\setcounter{section}{0}
\renewcommand{\thesection}{S\arabic{section}}
\setcounter{figure}{0}
\renewcommand{\thefigure}{S\arabic{figure}}
\setcounter{table}{0}
\renewcommand{\thetable}{S\arabic{table}}




\section{Evaluation methods}
\label{sec:evaluation_method}

\subsection{Spatial verification}

We calculate the critical success index (CSI), frequency bias (BIAS), false alarm ratio (FAR), and probability of detection (POD) using a mixed verification strategy. CR and TP are evaluated on the full gridded domain using neighborhood-based spatial verification, whereas GS is evaluated only on grid points selected by stations. Following NowcastNet \cite{zhang2023skilful}, the CR and TP forecast and reference fields are expanded with a max-pooling operator using a kernel size of 5, corresponding to a 5$\times$5\,km neighborhood at 1-km resolution. For a given threshold, an event is counted at a grid point if the maximum value within the corresponding neighborhood exceeds that threshold. This procedure reduces the double-penalty effect in deterministic high-resolution verification, in which a small position error can otherwise generate both a false alarm and a miss. For GS, the verification statistics are computed directly on the grid points selected by stations without neighborhood expansion.

For threshold-based evaluation, TP denotes true positives (hits) rather than the meteorological variable total precipitation used elsewhere in the manuscript; FP, FN, and TN denote false positives, false negatives, and true negatives, respectively. The CSI is defined as
\begin{equation}
    \text{CSI} = \frac{\text{TP}}{\text{TP} + \text{FP} + \text{FN}} .
\end{equation}
POD, FAR, and BIAS are defined as
\begin{equation}
    \text{POD} = \frac{\text{TP}}{\text{TP} + \text{FN}}, \quad
    \text{FAR} = \frac{\text{FP}}{\text{TP} + \text{FP}}, \quad
    \text{BIAS} = \frac{\text{TP} + \text{FP}}{\text{TP} + \text{FN}} .
\end{equation}
CSI, POD, and FAR range from 0 to 1, with higher CSI and POD and lower FAR indicating better performance; BIAS = 1 indicates an unbiased event frequency. We evaluate CR at 30, 40, and 50 dBZ; GS at 10.8, 13.9, and 17.2 m/s; and TP at 1, 20, and 50 mm.

\subsection{Baseline models}

For the persistence baseline, the most recent observed field is used as the forecast for all subsequent lead times. For each initialization, the field at $t=0$ is held fixed and treated as the prediction at every forecast hour.

For the extrapolation baseline, we use PySTEPS \cite{pulkkinen2019pysteps}. The motion field is estimated from consecutive radar composite reflectivity images using a Lucas--Kanade optical flow approach. This radar-derived motion field is then applied separately to the most recent composite reflectivity (CR), total precipitation (TP), and wind gust (GS) fields to generate deterministic semi-Lagrangian extrapolation forecasts for each variable. This setup is representative of extrapolation-based nowcasting. Because PySTEPS advects existing fields, it provides a strong extrapolation baseline but is not expected to generate new convective initiation without a precursor signal in the input.

\subsection{Additional verification metrics}

\begin{figure}[ht]
    \centering
    \includegraphics[width=\linewidth]{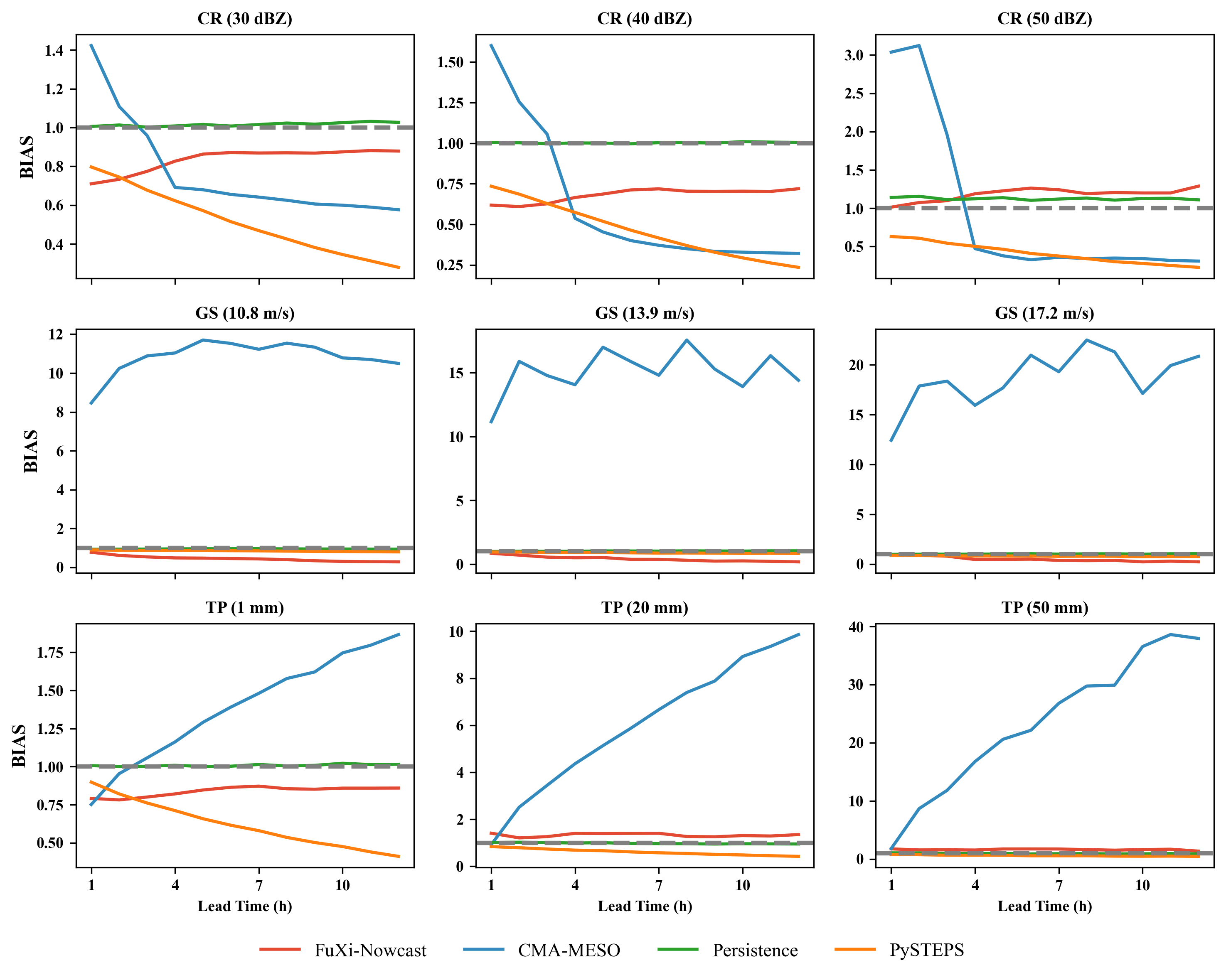}
    \caption{Frequency bias (BIAS) comparison among FuXi-Nowcast (red), CMA-MESO (blue), Persistence (green), and PySTEPS (orange) for 12\,h forecasts. Rows correspond to composite reflectivity (CR), wind gust (GS), and total precipitation (TP) from top to bottom. Columns show three intensity thresholds for each variable. The grey dashed line indicates the perfect BIAS value of 1.0. Values above 1 indicate overforecasting; values below 1 indicate underforecasting.}
    \label{fig:bias_comparison}
\end{figure}

\begin{figure}[ht]
    \centering
    \includegraphics[width=\linewidth]{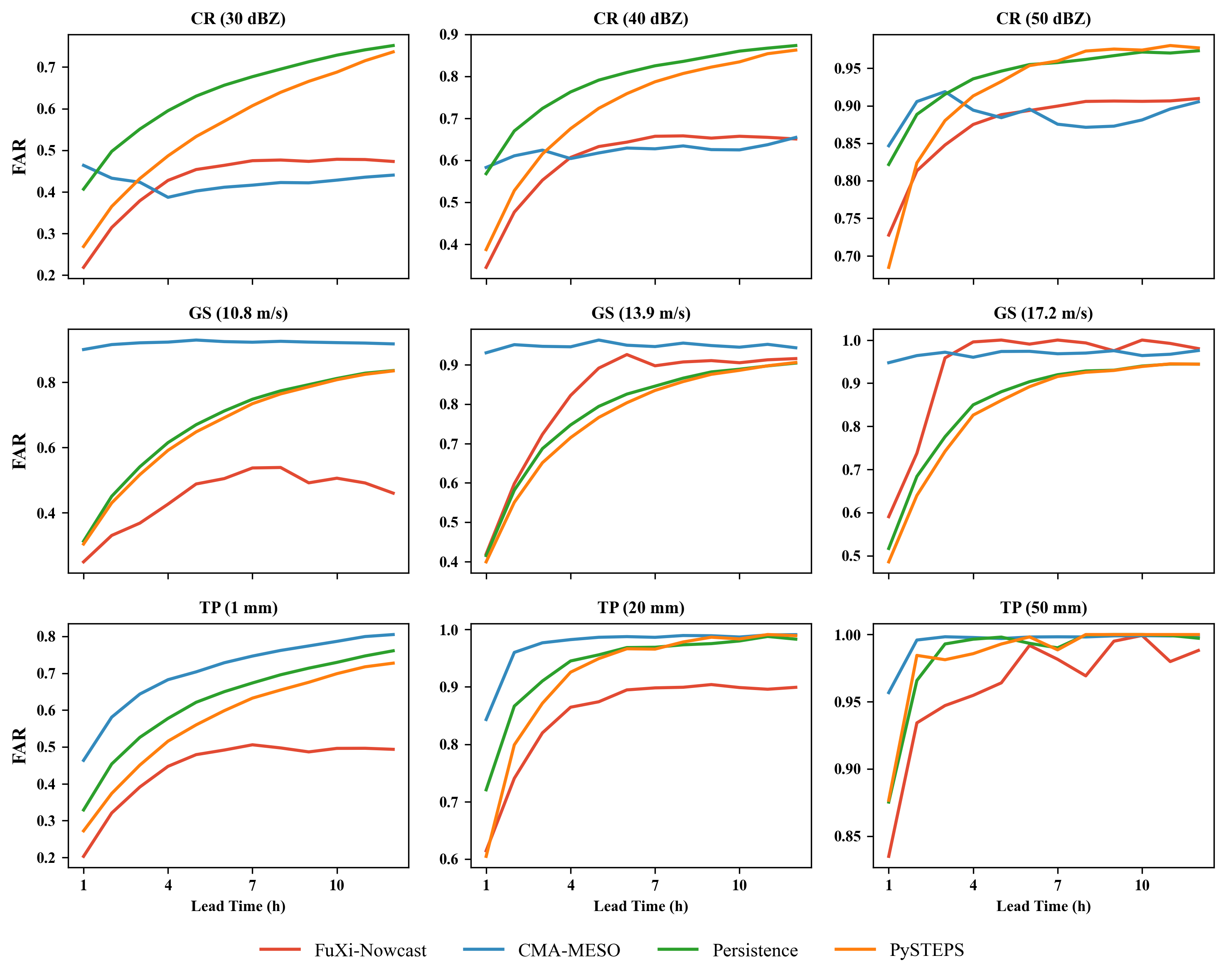}
    \caption{As in Supplementary Figure \ref{fig:bias_comparison}, but for false alarm ratio (FAR). Lower values indicate fewer false alarms. FAR = 0 corresponds to no false alarms.}
    \label{fig:far_comparison}
\end{figure}

\begin{figure}[ht]
    \centering
    \includegraphics[width=\linewidth]{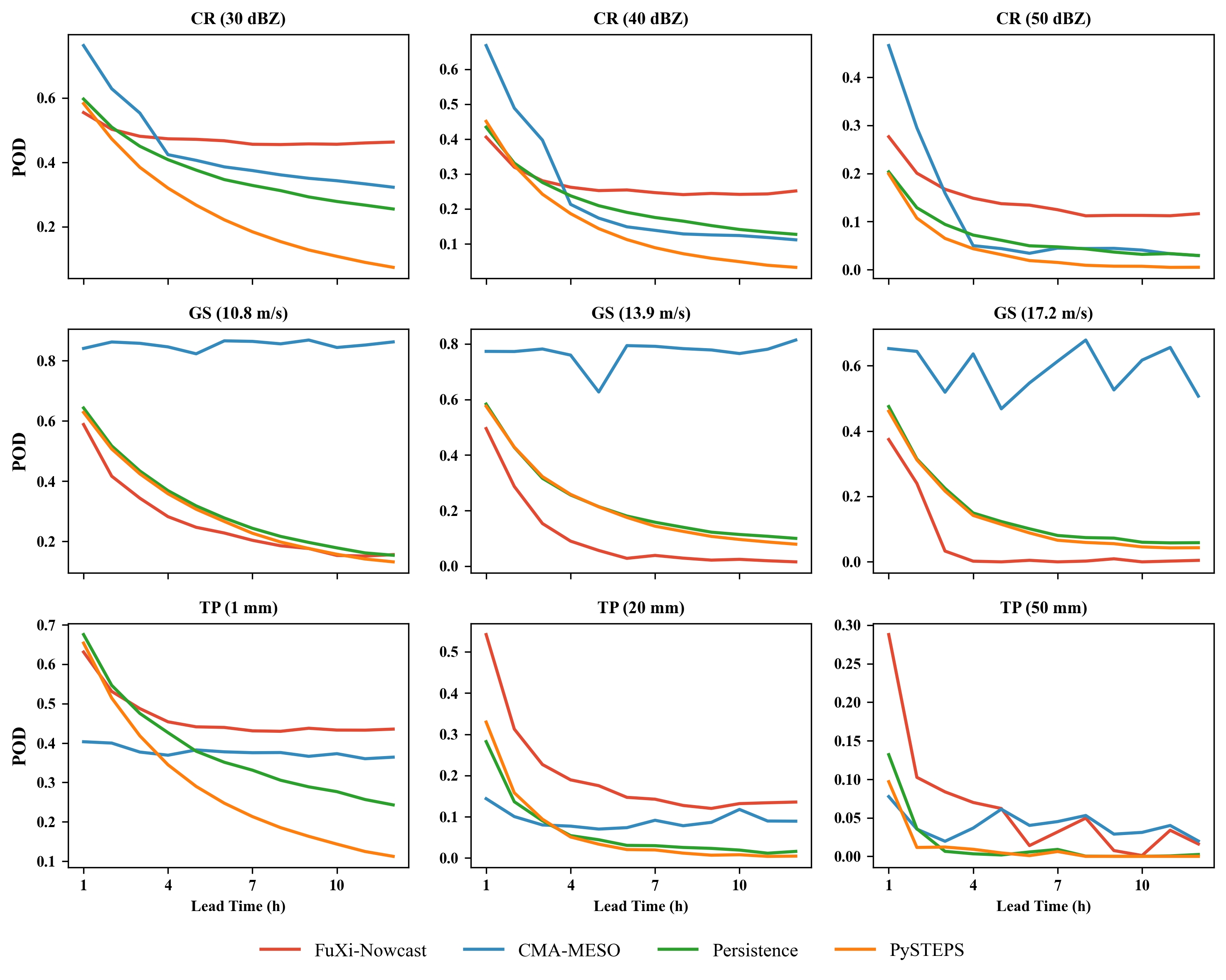}
    \caption{As in Supplementary Figure \ref{fig:bias_comparison}, but for probability of detection (POD). Higher values indicate better event detection. POD = 1 corresponds to all observed events being correctly predicted.}
    \label{fig:pod_comparison}
\end{figure}

The main text uses CSI as the primary verification metric. Here we report the corresponding BIAS, FAR, and POD for the same variables and thresholds (Supplementary Figures \ref{fig:bias_comparison}--\ref{fig:pod_comparison}) to further characterize model behavior.

BIAS (Supplementary Figure \ref{fig:bias_comparison}) quantifies systematic overprediction or underprediction. For FuXi-Nowcast, BIAS remains close to the ideal value of 1.0 for most variables and lead times, indicating limited systematic bias. The China Meteorological Administration Mesoscale Model (CMA-MESO) shows pronounced overforecasting for GS, especially at longer lead times, consistent with its low CSI in the main text. PySTEPS exhibits increasingly low BIAS with lead time, reflecting the progressive weakening of purely extrapolated echoes.

FAR (Supplementary Figure \ref{fig:far_comparison}) shows that, for CR and GS, FuXi-Nowcast generally has lower false-alarm rates than the other methods. In contrast, CMA-MESO has consistently high FAR values, indicating frequent overprediction of event occurrence and spatial extent.

POD (Supplementary Figure \ref{fig:pod_comparison}) shows that CMA-MESO can attain relatively high detection rates for CR at short lead times, but this is accompanied by substantially elevated false-alarm rates. FuXi-Nowcast shows a lower-FAR and lower-POD combination, which results in the highest overall CSI.

Overall, these additional metrics are consistent with the CSI results in the main text. FuXi-Nowcast achieves its advantage through a more balanced combination of detection and false alarms, whereas CMA-MESO often attains higher POD at the cost of excessive false alarms, and PySTEPS loses skill as the extrapolated echoes progressively weaken.

\FloatBarrier

\FloatBarrier

\FloatBarrier

\section{Data sources and preprocessing}

\begin{table}[ht]
    \centering
    \caption{Meteorological datasets used by FuXi-Nowcast, including original and target resolutions, variables and temporal coverage. The Role column indicates whether each variable is used as both input and output or only as input.}
    \scriptsize
    \setlength{\tabcolsep}{2pt}
    \begin{tabular}{@{}lccccc@{}}
        \toprule
        \textbf{Data Source} & \textbf{Role} & \begin{tabular}[c]{@{}c@{}}\textbf{Raw}\\\textbf{Resolution}\end{tabular} & \begin{tabular}[c]{@{}c@{}}\textbf{Processed}\\\textbf{Resolution}\end{tabular} & \textbf{Variables} & \textbf{Steps} \\
        \midrule
        FuXi-2.0 Forecasts & I &
        \begin{tabular}[c]{@{}c@{}}$721 \times 1440$\\(0.25$^\circ$)\end{tabular} &
        \begin{tabular}[c]{@{}c@{}}$768 \times 768$\\(0.01$^\circ$)\end{tabular} &
        \begin{tabular}[c]{@{}c@{}}65 upper-air variables\\(Z, T, U, V, Q)\\ 5 surface variables\\(T2M, U10M, V10M, MSL, TP)\end{tabular} & 3 \\
        Weather Stations & I \& O &
        \begin{tabular}[c]{@{}c@{}}--\\(0$^\circ$)\end{tabular} &
        \begin{tabular}[c]{@{}c@{}}$768 \times 768$\\(0.01$^\circ$)\end{tabular} &
        WS10M, GS & 3 \\
        Radar Reflectivity & I \& O &
        \begin{tabular}[c]{@{}c@{}}$1100 \times 1100$\\(0.01$^\circ$)\end{tabular} &
        \begin{tabular}[c]{@{}c@{}}$768 \times 768$\\(0.01$^\circ$)\end{tabular} &
        CR & 3 \\
        HRCLDAS & I \& O &
        \begin{tabular}[c]{@{}c@{}}$4500 \times 7000$\\(0.01$^\circ$)\end{tabular} &
        \begin{tabular}[c]{@{}c@{}}$768 \times 768$\\(0.01$^\circ$)\end{tabular} &
        Q2M, TP, U10M, V10M, T2M & 3 \\
        Land-Sea Mask & I &
        \begin{tabular}[c]{@{}c@{}}$721 \times 1440$\\(0.25$^\circ$)\end{tabular} &
        \begin{tabular}[c]{@{}c@{}}$768 \times 768$\\(0.01$^\circ$)\end{tabular} &
        Mask & 1 \\
        \bottomrule
    \end{tabular}
    \label{tab:all_data_si}
\end{table}
\FloatBarrier

Supplementary Table \ref{tab:all_data_si} summarizes the multi-source datasets used by FuXi-Nowcast. Below we provide the corresponding details on data sources, preprocessing, and normalization.

The ERA5 reanalysis dataset \cite{hersbach2020era5}, produced by the European Centre for Medium-Range Weather Forecasts (ECMWF), provides hourly global atmospheric data from January 1950 onward at a spatial resolution of about 31 km. To provide large-scale three-dimensional atmospheric fields during training, we use a subset of ERA5 at 0.25$^\circ$ spatial resolution and 1-hour temporal resolution. These data are linearly interpolated to 0.01$^\circ$ resolution and cropped to 29.01$^\circ$N--36.68$^\circ$N and 114.67$^\circ$E--122.34$^\circ$E ($768 \times 768$ grid points), matching the target domain of the other observations.

ERA5 has a latency of about five days and is therefore unsuitable for real-time inference. During inference, we instead use FuXi-2.0 forecasts \cite{chen2023fuxi,zhong2024fuxi}. FuXi-2.0 provides continuous 1-hour forecasts at 0.25$^\circ$ resolution and is initialized twice daily at 00 and 12 UTC. Its operational latency is about 8 hours. For example, forecasts initialized at 12 UTC become available at 20 UTC. For each FuXi-Nowcast run, we select the most recently available FuXi-2.0 cycle, extract the forecast hours valid at the nowcast target times, and use them as three-dimensional atmospheric fields. For example, a FuXi-Nowcast run initialized at 03 UTC on 3 July 2024 uses the FuXi-2.0 cycle from 12 UTC on 2 July, which has been available since 20 UTC on 2 July. Forecast lead times from T + 15 h to T + 20 h then provide the atmospheric fields valid from 03 to 08 UTC.

We use 70 meteorological variables from the FuXi-2.0 outputs. These include five upper-air variables: geopotential (Z), temperature (T), u component of wind (U), v component of wind (V), and specific humidity (Q). They are taken at 13 pressure levels: 50, 100, 150, 200, 250, 300, 400, 500, 600, 700, 850, 925, and 1000 hPa. We also use five surface variables: 2-meter temperature (T2M), 10-meter u wind component (U10M), 10-meter v wind component (V10M), mean sea-level pressure (MSL), and total precipitation (TP). All variables are interpolated to 1-km resolution and cropped to the same target domain.

In addition to the three-dimensional atmospheric fields, we incorporate multi-source ground observations. Surface wind data include 10-meter wind speed (WS10M) and wind gust (GS). Here GS is defined as the maximum instantaneous wind speed over a short duration within an hour \cite{suomi2018wind, chen2025study}. These data are collected from more than 1300 ground-based weather stations across East China. Station locations are shown in Supplementary Figure \ref{Landsea_MASK_stations}. WS10M is originally available at 6-minute resolution and is averaged to hourly means. GS is represented by hourly maxima. Both fields are then interpolated to the $768 \times 768$ grid using inverse distance weighting \cite{you2008comparison, zhao2022comparison}.

To provide direct precipitation information, we use radar composite reflectivity (CR) data at 1-km spatial resolution and 6-minute temporal resolution. These data cover 29.0$^\circ$N--40.0$^\circ$N and 113.0$^\circ$E--124.0$^\circ$E ($1100 \times 1100$ km). We resample the 6-minute radar data to hourly resolution by extracting instantaneous values at each hour. The fields are then cropped to the target domain.

The High-Resolution CMA Land Data Assimilation System (HRCLDAS) dataset \cite{han2020evaluation} integrates ground station observations, satellite data, and NWP model outputs. It provides high-quality atmospheric analyses for both model input and target. HRCLDAS provides hourly meteorological fields at 1-km spatial resolution over China and adjacent regions (15$^\circ$N--55$^\circ$N, 70$^\circ$E--140$^\circ$E). In this study, we use HRCLDAS data from April to September during 2019--2023 within the target domain. Five variables serve as the reference dataset for model training and evaluation: 2-meter specific humidity (Q2M), total precipitation (TP), 10-meter u wind component (U10M), 10-meter v wind component (V10M), and 2-meter temperature (T2M).

We also include a static land-sea mask from ERA5-Land \cite{munoz2021era5}. This mask distinguishes surface types and helps the model represent systematic differences in the evolution of precipitation and wind fields over land and ocean.

All input variables are normalized before training. For the 70 FuXi-2.0 atmospheric variables, TP is transformed using \(\log(1+\textrm{x})\). The remaining 69 variables use z-score normalization, \((\textrm{x} - \mu) / \sigma\). For the eight high-resolution local variables, WS, GS, CR, and TP are normalized by their respective maximum values, \(\textrm{x} / \max(\textrm{x})\). T2M, Q2M, U10M, and V10M use z-score normalization.

\section{Additional convective event cases}
\label{sec:additional_cases}

In addition to the dryline-triggered convective initiation (CI) case presented in the main text (16 June 2025), we examine additional cases from the 2024 evaluation period and the 2025 severe-weather season to show different aspects of FuXi-Nowcast's convective forecasting behavior, including scattered convective initiation, maintenance of organized convection over extended lead times, and the environmental conditions associated with a squall line and warm-sector convection. These cases are included to cover several typical modes of convective evolution discussed in this study. The 2024 cases are drawn from the main evaluation period, whereas the 2025 cases are presented as additional event analyses.

\subsection{Scattered convective initiation case: 12 June 2024 (03 UTC)}

\begin{figure}[ht]
    \centering
    \includegraphics[width=\linewidth]{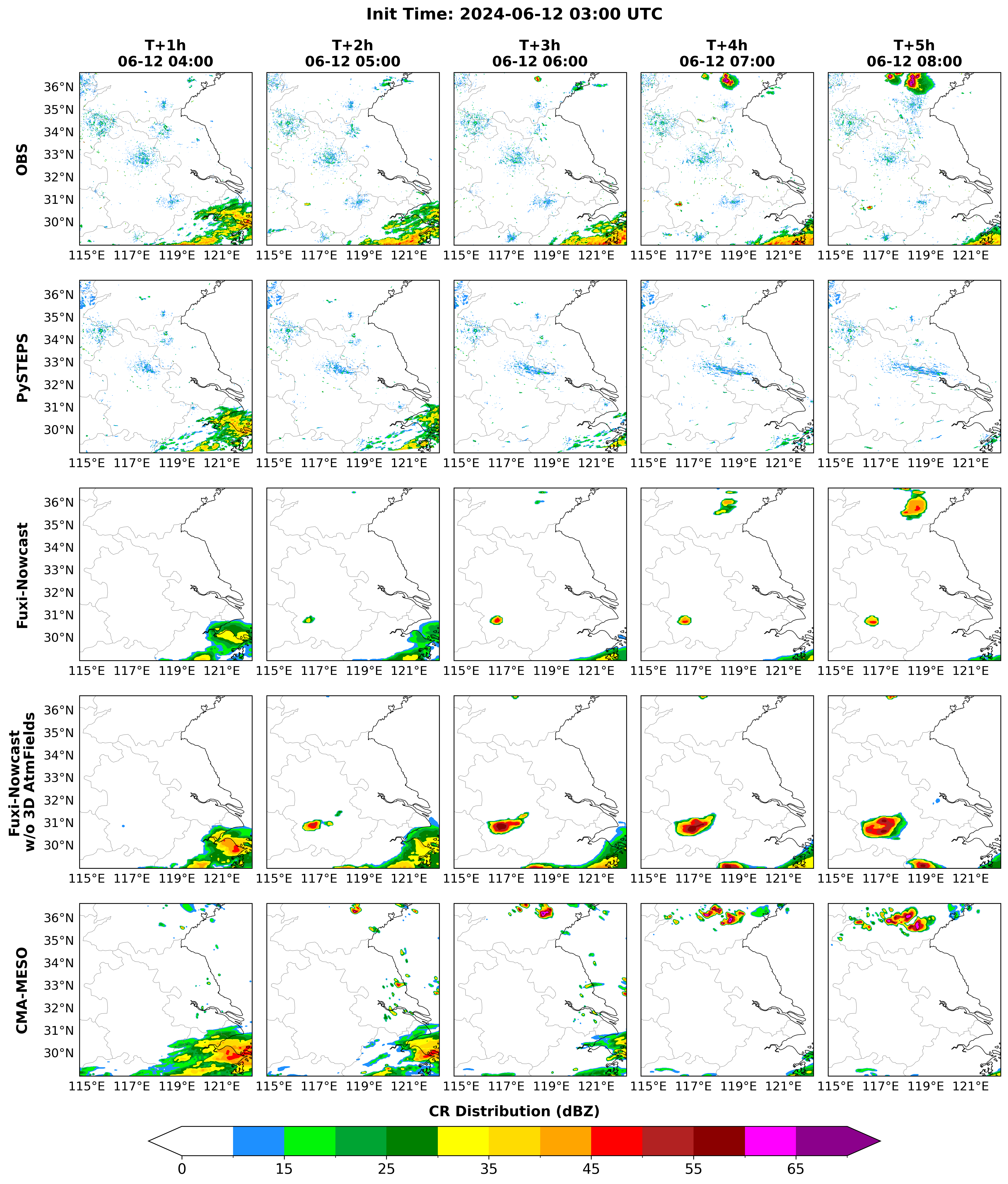}
    \caption{Composite radar reflectivity (dBZ) over East China for a scattered convective initiation case initialized at 03 UTC 12 June 2024. Rows from top to bottom show observations, PySTEPS forecasts, FuXi-Nowcast forecasts, the retrained FuXi-Nowcast variant without the three-dimensional atmospheric fields, and CMA-MESO forecasts, respectively. Columns from left to right correspond to forecast lead times of T + 1 to T + 5 h, valid from 04 to 08 UTC.}
    \label{fig:20240612T03}
\end{figure}

A scattered convective initiation event occurred over East China on 12 June 2024 (Supplementary Figure \ref{fig:20240612T03}).
Observed CR shows weak echoes in the northwestern part of the domain at T + 1 h (04 UTC), with isolated convective echoes developing and intensifying through T + 2--3 h (05--06 UTC) before dissipating by T + 4--5 h (07--08 UTC).
FuXi-Nowcast captures this evolution reasonably well: it predicts the initial appearance of scattered echoes in the northwest and their subsequent weakening.
The spatial distribution and intensity of the predicted reflectivity are generally consistent with the observations.

PySTEPS, by contrast, fails to capture CI and produces only a diffuse signal that rapidly fades.
CMA-MESO generates widespread moderate echoes across the northern part of East China from the outset that bear little resemblance to the observed isolated convective initiation in the northwest.
The figure also includes the retrained FuXi-Nowcast variant without the three-dimensional atmospheric fields for comparison.

\FloatBarrier

\subsection{Convective maintenance case: 21 June 2024}

\begin{figure}[ht]
    \centering
    \includegraphics[width=0.95\linewidth]{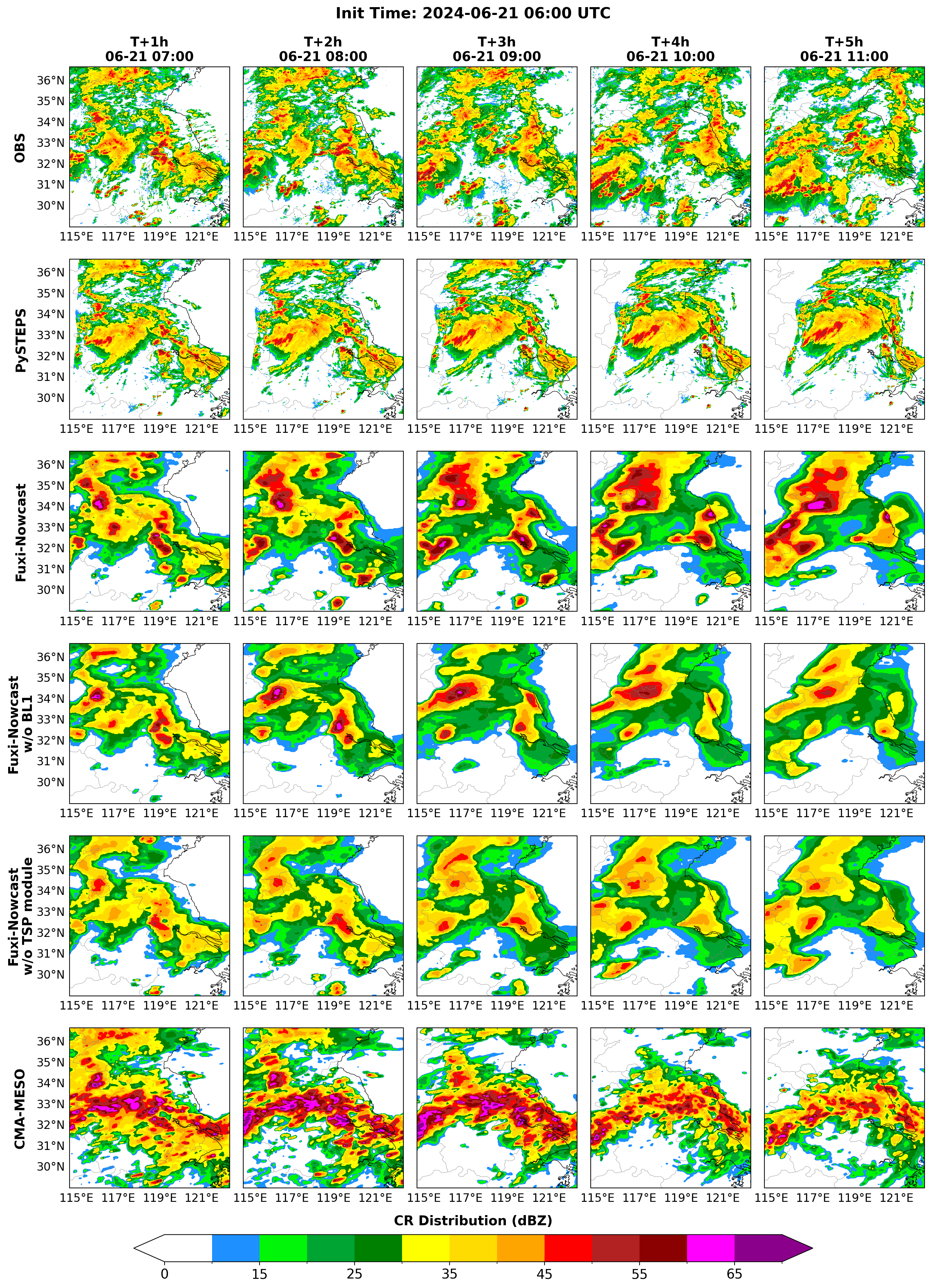}
    \caption{Composite radar reflectivity (dBZ) over East China for a convective maintenance case initialized at 06 UTC 21 June 2024. Rows from top to bottom show observations, PySTEPS forecasts, FuXi-Nowcast forecasts, two retrained ablation variants without the TSP module and without the Balanced L1 loss (BL1), and CMA-MESO forecasts, respectively. Columns from left to right correspond to forecast lead times of T + 1 to T + 5 h, valid from 07 to 11 UTC.}
    \label{fig:20240621T06}
\end{figure}

An extensive convective system affected East China on 21 June 2024 (Supplementary Figure \ref{fig:20240621T06}).
Observed CR at T + 1 h (07 UTC) shows a broad area of stratiform and embedded convective precipitation across the western and central parts of the domain.
Through T + 2--5 h (08--11 UTC), the system propagates eastward and intensifies, producing widespread reflectivity exceeding 45 dBZ with embedded convective echoes above 55 dBZ covering most of the province.

FuXi-Nowcast reproduces the large-scale organization and eastward propagation of this convective system.
The predicted reflectivity pattern captures both the stratiform background and the embedded convective maxima, while maintaining realistic intensity through the 5-hour forecast window.
PySTEPS preserves the initial echo pattern but progressively blurs and weakens the signal with increasing lead time, failing to maintain the convective structure beyond T + 2 h.
CMA-MESO overestimates the convective intensity and produces overly strong echoes in the southern part of the domain while underrepresenting the precipitation extent in the north.
The figure also includes two retrained ablation variants: one without the Thresholded Signal Pooling module (TSP module) and one without the Balanced L1 loss (BL1). Relative to the full FuXi-Nowcast, both variants lose the organization and intensity of the convective system more rapidly as lead time increases.

\FloatBarrier

\subsection{Squall line case: 27 June 2025}

A squall line event occurred over East China on 27 June 2025. Convection had already been triggered at the forecast initialization time (00 UTC). We therefore focus on the maintenance and structural evolution of the convective system. Comparison with observed composite reflectivity (CR) from 01 to 12 UTC shows that FuXi-Nowcast better reproduces the morphology and position of the squall line than CMA-MESO. However, it does not capture the observed weakening after 08 UTC and instead continues to intensify the system.

\begin{figure}[ht]
    \centering
    \includegraphics[width=\linewidth]{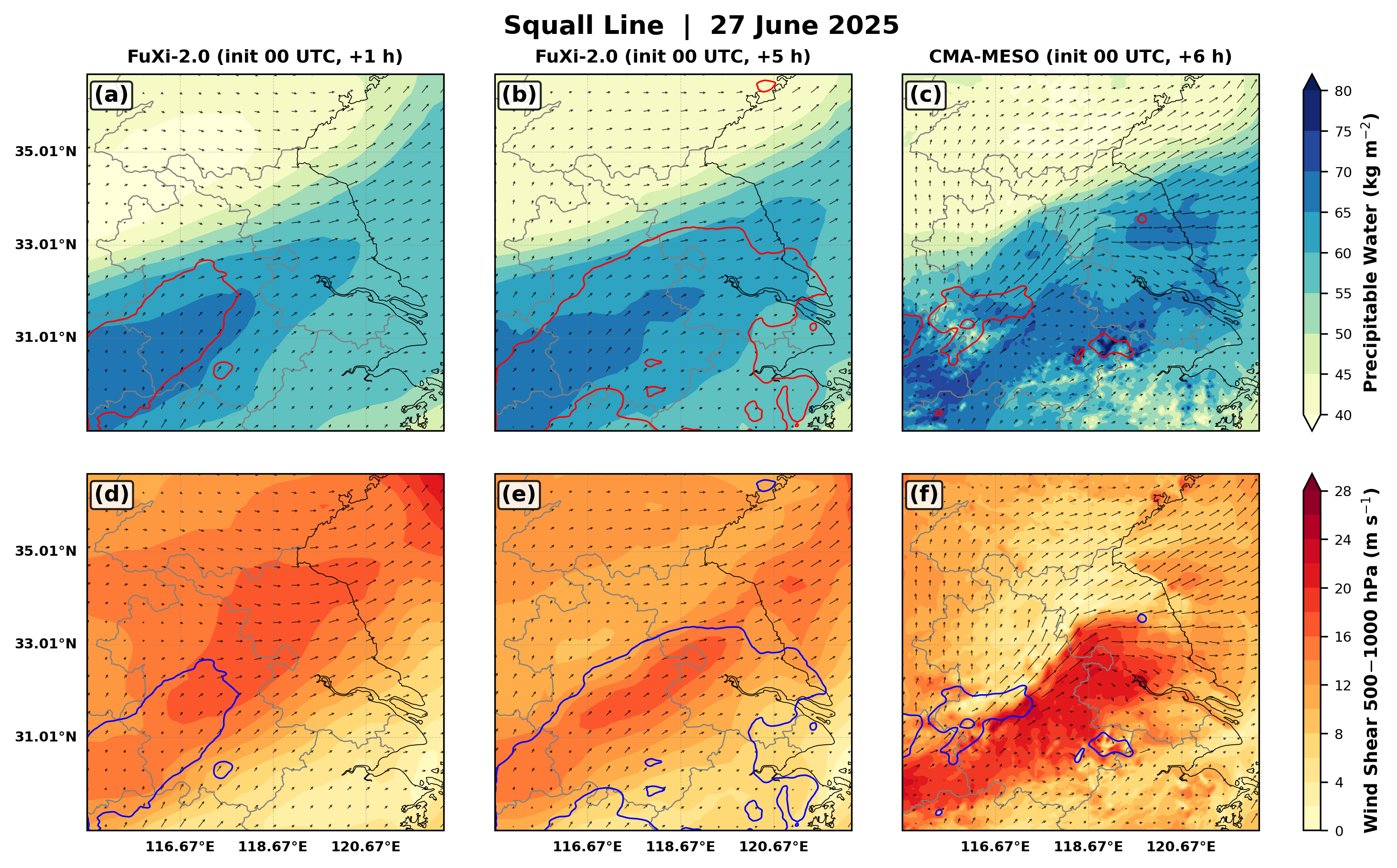}
    \caption{Environmental conditions diagnosed for the squall line case on 27 June 2025. Top row (a--c): total precipitable water (TPW; color shading, kg\,m$^{-2}$) with the FuXi-Nowcast 35 dBZ composite reflectivity contour (red) and 925 hPa wind vectors (arrows). Bottom row (d--f): 500--1000 hPa bulk wind shear magnitude (color shading, m\,s$^{-1}$) with the 35 dBZ contour (blue) and 925 hPa wind vectors. Columns from left to right: FuXi-2.0 three-dimensional atmospheric fields valid at T + 1 h (a, d) and T + 5 h (b, e) from the 00 UTC 27 June initialization, and CMA-MESO forecast fields valid at T + 6 h from the same initialization (c, f). All fields except composite reflectivity in columns 1--2 are from the FuXi-2.0 atmospheric fields; all fields in column 3 are CMA-MESO forecast variables.}
    \label{fig:case2_combined}
\end{figure}

To examine the different structural outcomes in the two models, we compare their environmental fields side by side (Supplementary Figure \ref{fig:case2_combined}). In the three-dimensional atmospheric fields at T + 1 h and T + 5 h (panels a, b, d, e), the predicted convective system (delineated by the 35 dBZ contour) is embedded within a region of total precipitable water (TPW) exceeding 70 kg\,m$^{-2}$ (panels a, b). It is also located within a region where the 500--1000 hPa bulk wind shear exceeds 14 m\,s$^{-1}$ (panels d, e). The moisture supply is consistent with sustained convective development. The strong deep-layer shear is also consistent with an environment favorable for storm organization and longevity, because it helps separate the updraft from the precipitation-driven downdraft. Together, these factors are consistent with conditions that support squall line formation and maintenance.

In contrast, CMA-MESO at T + 6 h (panels c, f) produces TPW as high as 75 kg\,m$^{-2}$ in the convective region (panel c). However, the 500--1000 hPa wind shear is only about 5 m\,s$^{-1}$ (panel f), which is less supportive of a coherent linear structure. The convection in CMA-MESO also appears in regions of high moisture content, but it lacks the same degree of dynamical organization seen in the atmospheric field diagnostics. This comparison indicates that wind-shear information may be important for representing this event.

\FloatBarrier

\section{Ablation studies}
\label{sec:ablation_studies}

This section presents detailed results from two sets of ablation experiments summarized in the main text: input variable ablations and model component ablations.
All experiments are evaluated using CSI, BIAS, FAR, and POD across CR, GS, and TP at multiple intensity thresholds over 12 h forecast horizons.
These experiments examines the contributions of atmospheric variables, model modules and loss terms to forecast performance.

Supplementary Table \ref{tab:ablation_settings} provides an overview of the retrained ablation configurations used in the manuscript, including the component ablations analyzed below and the variant without three-dimensional atmospheric fields shown in the convective initiation case studies.

\begin{table}[ht]
    \centering
    \caption{Overview of retrained ablation experiment settings for FuXi-Nowcast. Rows list the full model and the retrained ablation variants used in the manuscript. Columns indicate whether each component is retained or removed.}
    \scriptsize
    \setlength{\tabcolsep}{4pt}
    \begin{tabular}{@{}l*{8}{c}@{}}
        \toprule
        \textbf{Setting} & \textbf{\begin{tabular}[c]{@{}c@{}}3D \\ AtmFields\end{tabular}} & \textbf{TSP} & \textbf{ASF} & \textbf{BL1} & \textbf{Dice} & \textbf{Focal} & \textbf{LPIPS} & \textbf{SSIM} \\
        \midrule
        Full model & $\checkmark$ & $\checkmark$ & $\checkmark$ & $\checkmark$ & $\checkmark$ & $\checkmark$ & $\checkmark$ & $\checkmark$ \\
        w/o 3D atmospheric fields & $\times$ & $\checkmark$ & $\checkmark$ & $\checkmark$ & $\checkmark$ & $\checkmark$ & $\checkmark$ & $\checkmark$ \\
        w/o TSP module & $\checkmark$ & $\times$ & $\checkmark$ & $\checkmark$ & $\checkmark$ & $\checkmark$ & $\checkmark$ & $\checkmark$ \\
        w/o ASF module & $\checkmark$ & $\checkmark$ & $\times$ & $\checkmark$ & $\checkmark$ & $\checkmark$ & $\checkmark$ & $\checkmark$ \\
        w/o BL1 & $\checkmark$ & $\checkmark$ & $\checkmark$ & $\times$ & $\checkmark$ & $\checkmark$ & $\checkmark$ & $\checkmark$ \\
        w/o Dice & $\checkmark$ & $\checkmark$ & $\checkmark$ & $\checkmark$ & $\times$ & $\checkmark$ & $\checkmark$ & $\checkmark$ \\
        w/o Focal & $\checkmark$ & $\checkmark$ & $\checkmark$ & $\checkmark$ & $\checkmark$ & $\times$ & $\checkmark$ & $\checkmark$ \\
        w/o LPIPS & $\checkmark$ & $\checkmark$ & $\checkmark$ & $\checkmark$ & $\checkmark$ & $\checkmark$ & $\times$ & $\checkmark$ \\
        w/o SSIM & $\checkmark$ & $\checkmark$ & $\checkmark$ & $\checkmark$ & $\checkmark$ & $\checkmark$ & $\checkmark$ & $\times$ \\
        \bottomrule
    \end{tabular}
    \label{tab:ablation_settings}
\end{table}
\FloatBarrier

\subsection{Input variable ablation}

\begin{figure}[ht]
    \centering
    \includegraphics[width=\linewidth]{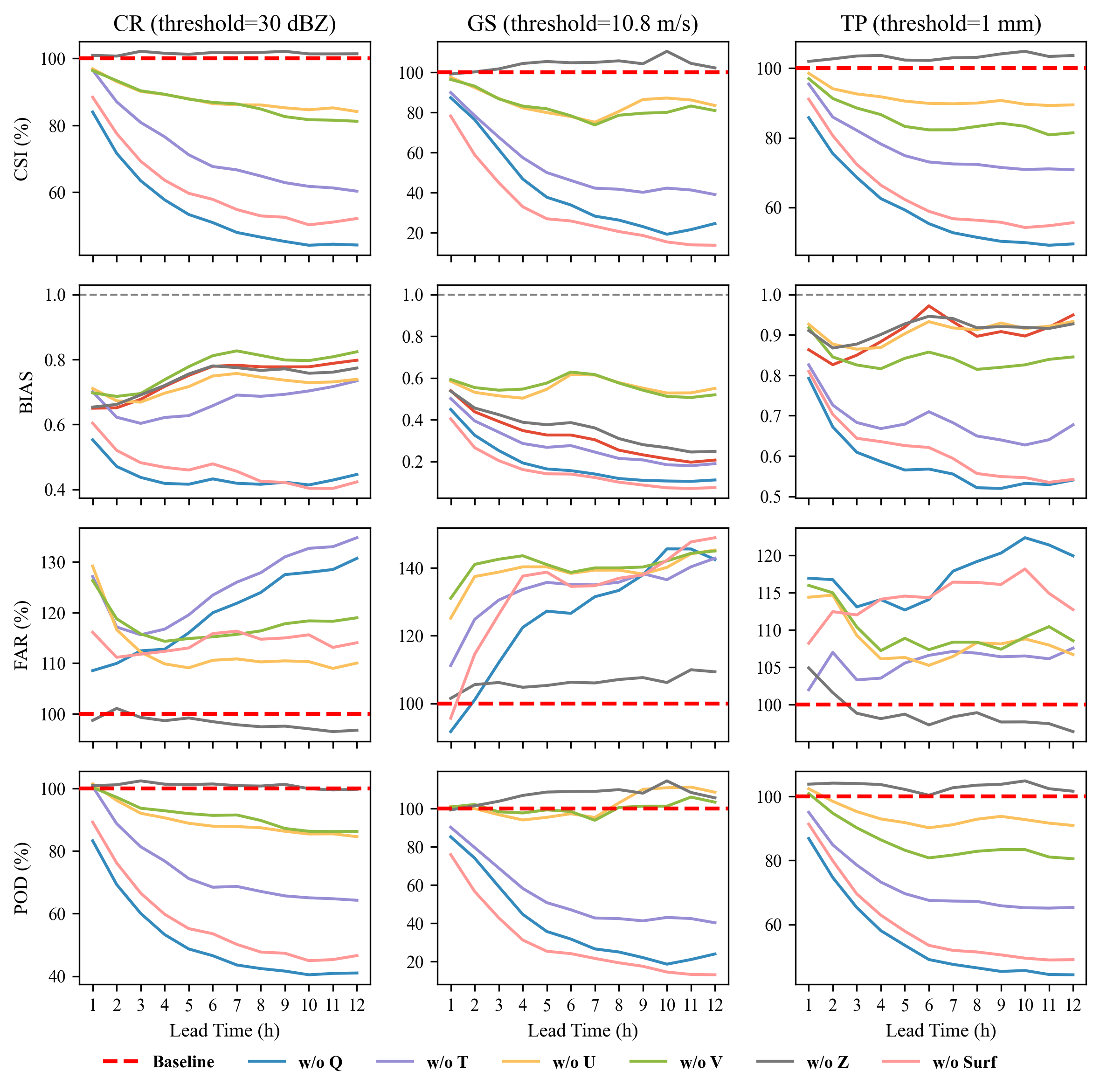}
    \caption{Ablation study showing relative verification changes when individual variable groups are removed from the input. Columns correspond to composite reflectivity (CR, threshold = 30 dBZ), wind gust (GS, threshold = 10.8 m/s), and total precipitation (TP, threshold = 1 mm), respectively. Rows show relative CSI (\%), BIAS, relative FAR (\%), and relative POD (\%) from top to bottom. The red dashed line in the relative CSI, relative FAR, and relative POD panels denotes the baseline performance (100\%), while the grey dashed line in the BIAS panels indicates the perfect value of 1.0. Each colored line represents a different ablation experiment: removing specific humidity (w/o Q), temperature (w/o T), the u component of wind (w/o U), the v component of wind (w/o V), geopotential (w/o Z), or the five surface atmospheric variables T2M, U10M, V10M, MSL, and TP (w/o Surf).}
    \label{fig:ablation_combined}
\end{figure}

To quantify the contribution of each input variable group, we zero out its channels and retrain the model from scratch using identical hyperparameters and training schedules as the baseline (AdamW optimizer, 30,000 iterations, same learning rate warm-up and cosine decay). This experiment is intended as a model sensitivity analysis under the current input design rather than as a strict physical attribution analysis.
Six groups are tested: specific humidity (Q), temperature (T), the u component of wind (U), the v component of wind (V), geopotential (Z), and all surface variables (Surf).
Supplementary Figure \ref{fig:ablation_combined} shows relative CSI, BIAS, relative FAR, and relative POD as a function of lead time for each ablation experiment.

Removing Q causes the most severe degradation for CR (CSI dropping to $\sim$40\% at 12 h) and TP (CSI dropping to $\sim$50\%), while surface variables are most critical for GS (CSI dropping to $\sim$20\%).
Temperature and wind components show moderate contributions, and geopotential has the smallest impact, suggesting partial redundancy with other variables.

\FloatBarrier

\subsection{Model component ablation}

\begin{figure}[ht]
    \centering
    \includegraphics[width=\linewidth]{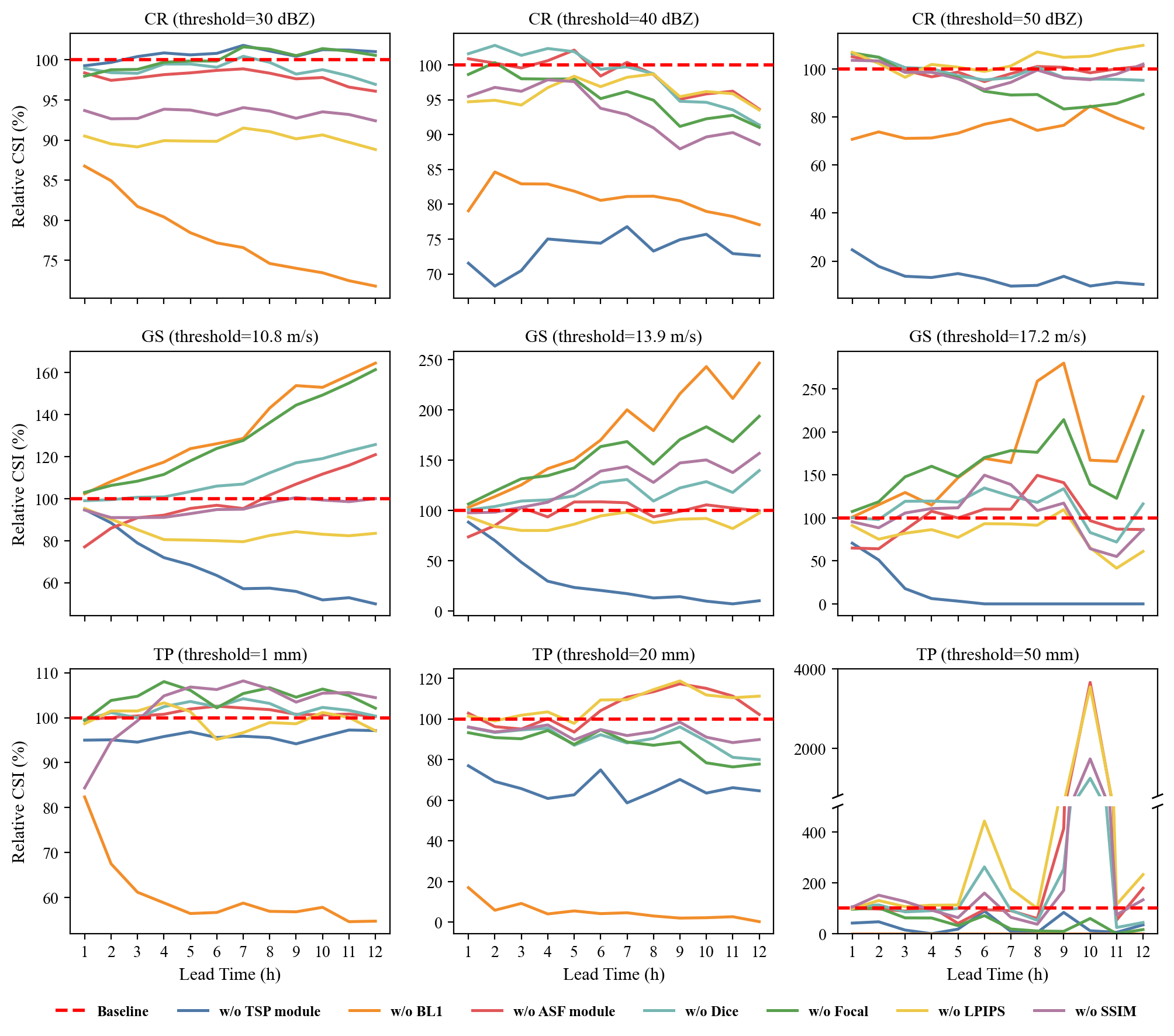}
    \caption{Ablation study on model components, showing relative CSI (\%) as a function of lead time. Rows correspond to composite reflectivity (CR; 30, 40, 50 dBZ), wind gust (GS; 10.8, 13.9, 17.2 m/s), and total precipitation (TP; 1, 20, 50 mm) from top to bottom. The red dashed line denotes the baseline performance (100\%). The TP 50 mm panel uses a broken y-axis to show several extreme relative values. Each colored line represents a different ablation experiment: removing the Thresholded Signal Pooling module (w/o TSP module), the Balanced L1 loss (w/o BL1), the Adaptive Signal Fusion module (w/o ASF module), the Dice loss (w/o Dice), the Focal loss (w/o Focal), the LPIPS loss (w/o LPIPS), or the SSIM loss (w/o SSIM).}
    \label{fig:method_ablation_csi}
\end{figure}

We further evaluate the contribution of key model components by individually removing the Thresholded Signal Pooling module (TSP module), which reinforces strong convective signals, the Adaptive Signal Fusion module (ASF module), which mixes the enhanced signals back into the original input, and five loss terms, measuring the relative CSI change across CR, GS, and TP at multiple thresholds (Supplementary Figure \ref{fig:method_ablation_csi}).
When a loss component is removed, the weights of all remaining loss terms are kept unchanged; each ablation model is retrained from scratch with identical hyperparameters and training schedules as the baseline.

The TSP module and the Balanced L1 loss are the two most critical components.
Removing the TSP module reduces CSI to $\sim$70\% for CR at 30 dBZ and below 20\% at 50 dBZ; for GS the degradation is even more severe, with CSI dropping near zero at 17.2 m/s.
Removing Balanced L1 causes near-complete failure for heavy precipitation (TP at 20 mm approaching zero) and also degrades CR prediction.
The remaining components---the Focal loss, Dice loss, learned perceptual image patch similarity (LPIPS) loss, structural similarity index measure (SSIM) loss, and the Adaptive Signal Fusion module (ASF module)---each contribute moderate improvements of 5--15\% CSI across different variables and thresholds.

These results indicate that the two dominant components act on different aspects of forecast quality. The TSP module is most important for retaining strong convective signatures, while the Balanced L1 loss is most important for rare heavy-precipitation events.

\FloatBarrier

\section{Data and station network}
\label{sec:data_figures}

\begin{figure}[ht]
    \centering
    \includegraphics[width=\linewidth]{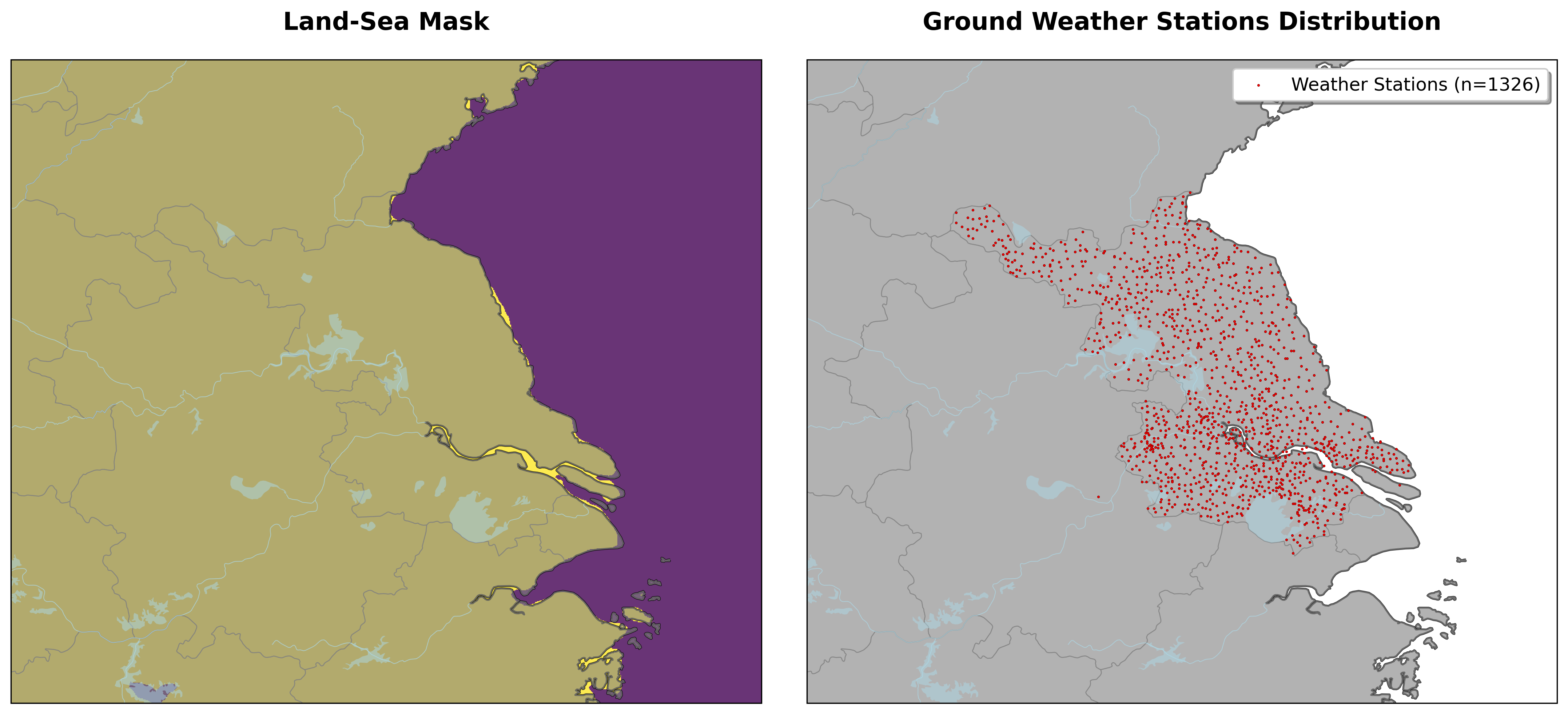}
    \caption{Spatial distribution of ground-based weather stations across East China. The station network consists of fixed station locations, with occasional data gaps at some stations during certain time periods.}
    \label{Landsea_MASK_stations}
\end{figure}

This section provides additional details on the observational data used in this study.
Supplementary Figure \ref{Landsea_MASK_stations} shows the gridded ground-weather-station distribution across East China.
The network comprises more than 1300 fixed stations that provide hourly observations of wind speed, wind gusts, temperature, and humidity.
Occasional data gaps occur at individual stations during certain time periods, but do not substantially affect the overall spatial coverage.
The station mask used in evaluation is identical to this gridded station distribution and is used only to select GS verification grid points, whereas CR and TP are evaluated on the full gridded domain.

\FloatBarrier

\begin{figure}[ht]
    \centering
    \includegraphics[width=\linewidth]{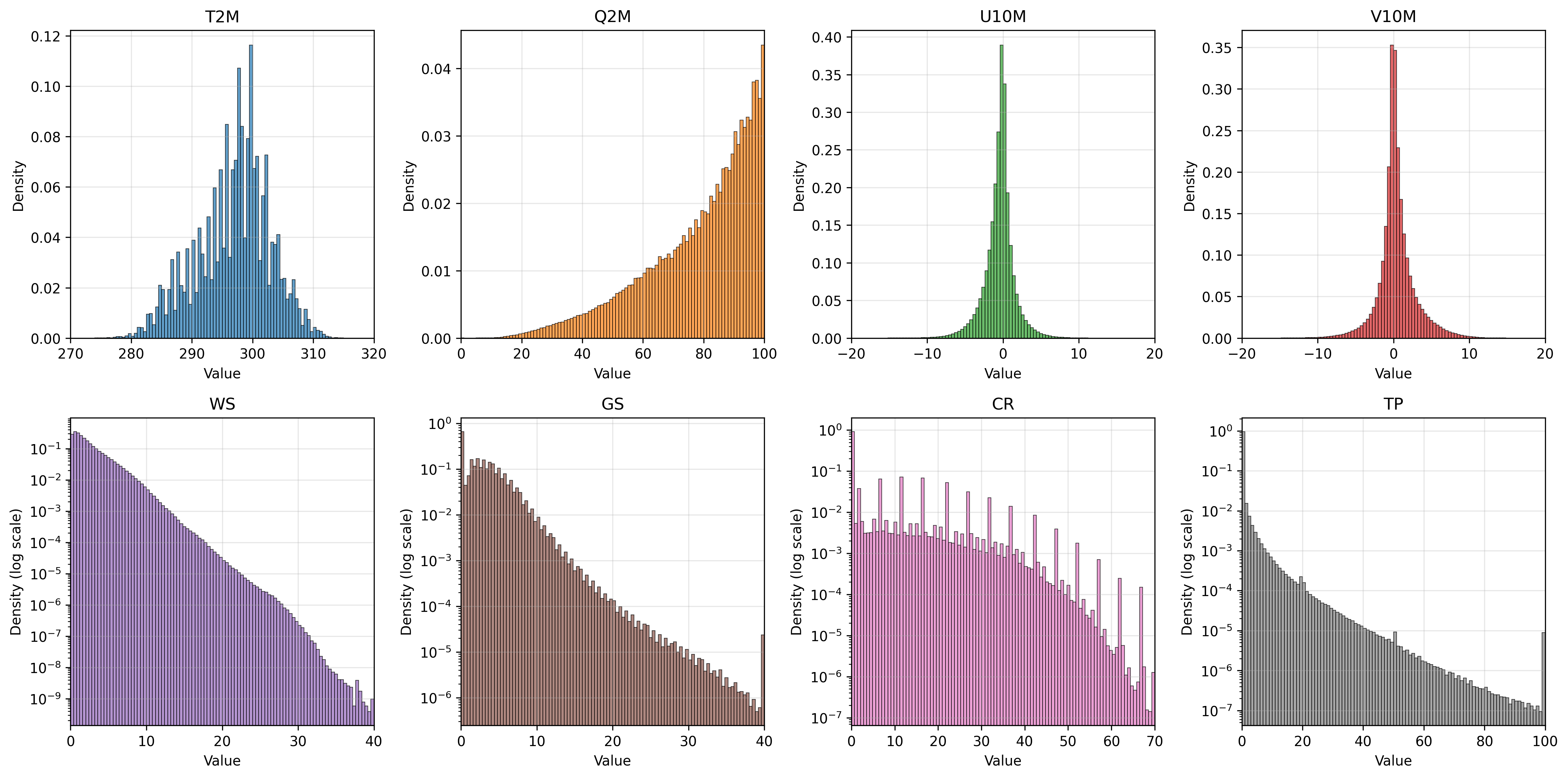}
    \caption{Statistical distributions of the eight target forecast variables before sample filtering. The top row shows 2-meter temperature (T2M), 2-meter specific humidity (Q2M), 10-meter u wind component (U10M), and 10-meter v wind component (V10M), from left to right. The bottom row shows wind speed (WS), wind gusts (GS), composite reflectivity (CR), and total precipitation (TP), from left to right. The y-axis is plotted on a linear scale for the top row and on a logarithmic scale for the bottom row.}
    \label{fig:variable_distributions}
\end{figure}

\begin{figure}[ht]
    \centering
    \includegraphics[width=\linewidth]{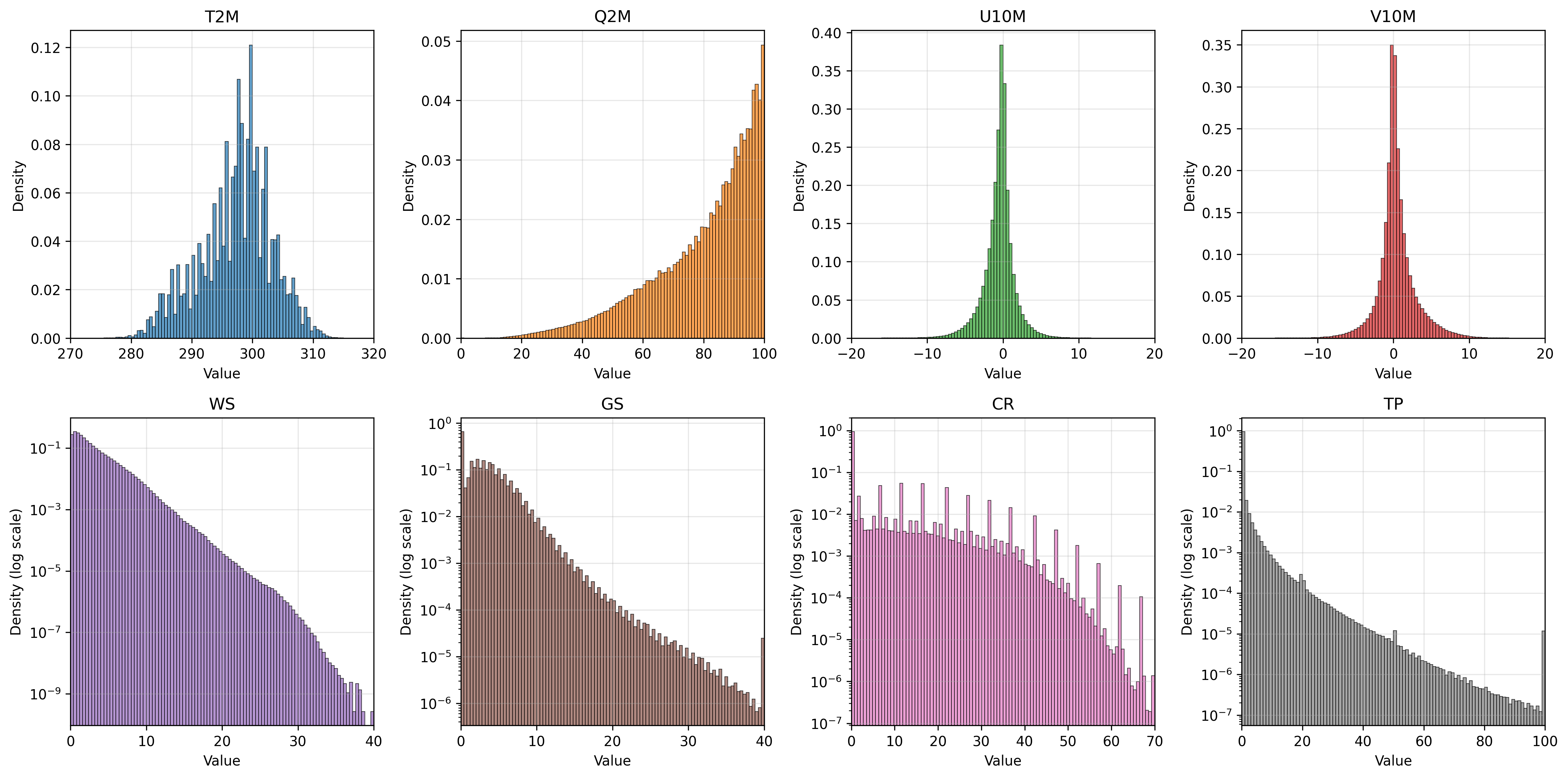}
    \caption{As in Supplementary Figure \ref{fig:variable_distributions}, but after applying the sample-filtering procedure that retains only cases with significant precipitation or wind-gust events in the training dataset.}
    \label{fig:variables_distribution_filter}
\end{figure}

Supplementary Figures \ref{fig:variable_distributions} and \ref{fig:variables_distribution_filter} show the statistical distributions of the eight target forecast variables before and after the date filtering procedure described in the main text. This filtering retains 13,672 out of 18,546 total samples (73.7\%). It increases the proportion of high-impact weather cases in the training dataset while preserving a sufficiently broad sample distribution. No filtering is applied to the testing dataset, so it continues to represent the full range of weather conditions. As a result, the filtered distributions show higher probability mass in the tails for TP, CR, and GS, while the near-Gaussian variables (T2M, Q2M, U10M, V10M) remain largely unchanged.

\FloatBarrier

\FloatBarrier

\section{Diagnostic variables}
\label{sec:diagnostic_variables}

This section provides the definitions and derivations of the standard meteorological diagnostic variables used in the physical mechanism analysis.

\subsection{Wind field divergence}

Wind field divergence quantifies the horizontal spreading or convergence of air parcels and is used here as a diagnostic indicator of low-level forcing:

\begin{equation}
\label{eq:divergence}
    \textrm{D} = \nabla \cdot \mathbf{U} = \frac{\partial \textrm{u}}{\partial \textrm{x}} + \frac{\partial \textrm{v}}{\partial \textrm{y}}
\end{equation}

where $\textrm{D}$ is horizontal wind divergence (s$^{-1}$), $\mathbf{U} = (\textrm{u},\textrm{v})$ is the horizontal wind vector, $\textrm{u}$ is the zonal wind component (m\,s$^{-1}$), $\textrm{v}$ is the meridional wind component (m\,s$^{-1}$), and $\textrm{x}$ and $\textrm{y}$ are the eastward and northward horizontal coordinates (m), respectively. Negative values of $\textrm{D}$ indicate convergence.

\subsection{Relative humidity}

Relative humidity is defined as the ratio of water-vapor pressure to saturation vapor pressure:

\begin{equation}
\textrm{RH} = \frac{\textrm{e}}{\textrm{e}_\textrm{s}(\textrm{T})}
\end{equation}

where $\textrm{RH}$ is relative humidity (dimensionless, 0--1), $\textrm{e}$ is water-vapor pressure (Pa), and $\textrm{e}_\textrm{s}(\textrm{T})$ is saturation vapor pressure (Pa) at air temperature $T$.
In this study, $e$ is obtained from pressure $\textrm{p}$ and specific humidity $\textrm{q}$ as:

\begin{equation}
\textrm{e} = \frac{\textrm{q} \textrm{p}}{\varepsilon + (1-\varepsilon)\textrm{q}}
\end{equation}

where $\textrm{q}$ is specific humidity (kg\,kg$^{-1}$), $\textrm{p}$ is air pressure (Pa), and $\varepsilon = 0.622$ is the ratio of the gas constants of dry air and water vapor.
The saturation vapor pressure is calculated with the Bolton approximation:

\begin{equation}
\textrm{e}_\textrm{s}(\textrm{T}) = 611.2 \exp\left[\frac{17.67(\textrm{T}-273.15)}{\textrm{T}-29.65}\right]
\end{equation}

where $\textrm{T}$ is air temperature (K). In all formulas below, $\textrm{RH}$ is treated as a dimensionless quantity between 0 and 1. If needed, relative humidity in percent is obtained as $100 \times \textrm{RH}$.

\subsection{Dew point temperature}

The dew point temperature is the temperature to which air must be cooled, at constant pressure, to reach saturation. It is computed from water-vapor pressure in Pa using the Bolton approximation:

\begin{equation}
\label{eq:dewpoint_temperature}
    \textrm{T}_\textrm{d} = \frac{243.5 \ln(\textrm{e} / 611.2)}{17.67 - \ln(\textrm{e} / 611.2)}
\end{equation}

where $\textrm{T}_\textrm{d}$ is dew point temperature ($^\circ$C) and $\textrm{e}$ is water-vapor pressure (Pa). In practice, $\textrm{e} = \textrm{RH} \cdot \textrm{e}_\textrm{s}(\textrm{T}_\textrm{c})$, where $RH$ is relative humidity (dimensionless, 0--1), $T_c$ is air temperature ($^\circ$C), and $\textrm{e}_\textrm{s}(\textrm{T}_\textrm{c})$ is the saturation vapor pressure (Pa) at $\textrm{T}_\textrm{c}$:

\begin{equation}
\textrm{e}_\textrm{s}(\textrm{T}_\textrm{c}) = 611.2 \exp\left(\frac{17.67 \textrm{T}_\textrm{c}}{\textrm{T}_\textrm{c} + 243.5}\right)
\end{equation}

\subsection{Precipitable water}

Total precipitable water measures the vertically integrated atmospheric water vapor in a column:

\begin{equation}
\label{eq:precipitable_water}
\textrm{TPW} = \frac{1}{ \rho_\textrm{w} \textrm{g}} \int_{\textrm{p}_{\mathrm{top}}}^{\textrm{p}_{\mathrm{bottom}}} \textrm{q} \, \textrm{dp} 
\end{equation}
where $\textrm{TPW}$ is total precipitable water (mm), $\rho_\textrm{w}$ is the density of liquid water (approximately 1000 kg\,m$^{-3}$), $\textrm{g}$ is gravitational acceleration (approximately 9.81 m\,s$^{-2}$), $\textrm{q}$ is specific humidity (kg\,kg$^{-1}$), and $\textrm{p}_{\mathrm{top}}$ and $\textrm{p}_{\mathrm{bottom}}$ are the pressures at the top and bottom of the integration layer (Pa), respectively.
Under this formulation, 1 mm of $\textrm{TPW}$ is equivalent to 1 kg\,m$^{-2}$ of vertically integrated water vapor.

\subsection{Equivalent potential temperature}

Equivalent potential temperature is computed following Bolton (1980) \cite{bolton1980computation}. First, the lifting condensation level (LCL) temperature is computed as:

\begin{equation}
\label{eq:lcl_temperature}
\textrm{T}_\textrm{L} = \frac{1}{\frac{1}{\textrm{T}_\textrm{D}-56}+\frac{\ln(\textrm{T}_\textrm{K}/\textrm{T}_\textrm{D})}{800}}+56
\end{equation}
where temperatures in the Bolton formula are in K, pressure is in Pa, and mixing ratio is in kg kg$^{-1}$.

The potential temperature at the LCL is then computed as:

\begin{equation}
\label{eq:potential_temperature_lcl}
\theta_\textrm{DL} = \textrm{T}_\textrm{K}\left(\frac{100000}{\textrm{p}-\textrm{e}}\right)^\kappa \left(\frac{\textrm{T}_\textrm{K}}{\textrm{T}_\textrm{L}}\right)^{0.28\textrm{r}}
\end{equation}

The equivalent potential temperature is:

\begin{equation}
\label{eq:equivalent_potential_temperature}
\theta_\textrm{E} = \theta_\textrm{DL}\exp\left[\left(\frac{3036}{\textrm{T}_\textrm{L}} -1.78\right)\textrm{r}(1+0.448\textrm{r})\right]
\end{equation}

where $\textrm{T}_\textrm{L}$ is the LCL temperature (K), $\textrm{T}_\textrm{K}$ is the parcel temperature (K), $\textrm{T}_\textrm{D}$ is the dew point temperature (K), $\theta_\textrm{DL}$ is the potential temperature at the LCL (K), $\theta_\textrm{E}$ is the equivalent potential temperature (K), $\textrm{r}$ is the saturation mixing ratio (kg\,kg$^{-1}$), $\textrm{p}$ is pressure (Pa), $\textrm{e}$ is the saturation vapor pressure at dew point (Pa), and $\kappa \approx 0.286$ is the Poisson constant for dry air.

\subsection{Convective available potential energy}

CAPE quantifies the vertically integrated buoyant energy available to a lifted air parcel:

\begin{equation}
\label{eq:cape}
\text{CAPE} = \int_{\textrm{p}_{\textrm{LFC}}}^{\textrm{p}_{\text{EL}}} \textrm{R}_\textrm{d} \left( \textrm{T}_{\textrm{v},\text{parcel}} - \textrm{T}_{\textrm{v},\text{env}} \right) \textrm{d}\ln \textrm{p}
\end{equation}
where $\textrm{T}_{\textrm{v},\text{parcel}}$ is the parcel virtual temperature (K), $\textrm{T}_{\textrm{v},\text{env}}$ is the environmental virtual temperature (K), $\textrm{R}_\textrm{d}$ is the gas constant for dry air (287 J\,kg$^{-1}$\,K$^{-1}$), $\textrm{p}$ is pressure (Pa), $\textrm{p}_{\text{LFC}}$ is the pressure at the level of free convection (Pa), and $\textrm{p}_{\text{EL}}$ is the pressure at the equilibrium level (Pa).
Because pressure decreases with height, the pressure-coordinate integral is written from $p_{\text{EL}}$ to $p_{\text{LFC}}$ so that positive buoyancy yields positive CAPE.
The integral is evaluated only over the positively buoyant layer, and CAPE is expressed in J\,kg$^{-1}$.

\clearpage
\bibliography{refs}

\end{document}